\documentclass[aps,superscriptaddress,prc,nofootinbib,showpacs,showkeys]{revtex4-2}
\usepackage{graphicx}
\usepackage{amssymb}
\usepackage{amsmath}
\usepackage{amsthm}
\usepackage{color}
\usepackage[normalem]{ulem}
\usepackage{slashed}
\usepackage{mathrsfs}
\usepackage{hyperref}
\usepackage[mathscr]{eucal}
\usepackage{bigints}
\hypersetup{
	colorlinks=true,        
	linkcolor=blue,         
	citecolor=blue,         
}

\allowdisplaybreaks[4]

\begin{document}

\title{Correlations of dihadron polarization in central, peripheral and ultraperipheral heavy-ion collisions}
\author{Xiaowen Li}
\affiliation{Key Laboratory of Particle Physics and Particle Irradiation (MOE), Institute of frontier and interdisciplinary science, Shandong University, Qingdao, Shandong 266237, China}
	
\author{Zhao-Xuan Chen}
\affiliation{Key Laboratory of Particle Physics and Particle Irradiation (MOE), Institute of frontier and interdisciplinary science, Shandong University, Qingdao, Shandong 266237, China}
	
\author{Shanshan Cao}
\email{shanshan.cao@sdu.edu.cn}
\affiliation{Key Laboratory of Particle Physics and Particle Irradiation (MOE), Institute of frontier and interdisciplinary science, Shandong University, Qingdao, Shandong 266237, China}
	
\author{Shu-Yi Wei}
\email{shuyi@sdu.edu.cn}
\affiliation{Key Laboratory of Particle Physics and Particle Irradiation (MOE), Institute of frontier and interdisciplinary science, Shandong University, Qingdao, Shandong 266237, China}

\begin{abstract}
While jet quenching in relativistic heavy-ion collisions has been extensively studied over decades, the polarization of quenched hadrons has rarely been discussed. It has recently been proposed that the correlations of dihadron polarization in $e^+e^-$ and $pp$ collisions provide a novel probe of the longitudinal spin transfer from hard partons to hadrons without requiring the colliding beams to be polarized. To support realistic experimental measurement of dihadron polarization with sufficient luminosity, we extend the aforementioned study to relativistic heavy-ion collisions by convoluting the vacuum fragmentation of partons with their energy loss inside the quark-gluon plasma (QGP). We find that while the correlation functions of $\Lambda$-$\Lambda$ (or $\Lambda$-$\bar{\Lambda}$) polarization in peripheral collisions is consistent with those in $pp$ collisions, clear enhancement can be seen in central collisions. These correlation functions appear sensitive to different assumptions in the DSV parameterization of parton fragmentation functions, and therefore could place additional constraints on the spin-dependent fragmentation functions of quarks and gluons. The correlation of dihadron polarization has also been explored in ultraperipheral heavy-ion collisions, which provides a cleaner probe of fragmentation functions of quarks produced by energetic photon-photon and photon-pomeron interactions.
\end{abstract}	

\maketitle	

\section{introduction}
\label{section1}

A jet is a spray of collimated hadrons that emanate from a high-energy (hard) parton produced in energetic leptonic or nuclear collisions. The cross section of hadron production can be factorized into that of parton production and fragmentation functions (FFs) of these partons, the latter of which describe the distributions of momentum fraction and spin inherited by a hadron from its parent parton. Fragmentation functions serve as a crucial bridge connecting partonic quantities that are perturbatively calculable and hadronic observables from experiments \cite{Metz:2016swz,Chen:2023kqw}. To fully address the production of hadrons with different spins, a bunch of FFs need to be introduced at the leading twist~\cite{Mulders:1995dh,Boer:1997mf,Bacchetta:2006tn,Wei:2013csa,Wei:2014pma}, while there are even more at higher twists~\cite{Levelt:1994np,Mulders:1995dh,Boer:1997mf,Boer:1997qn,Bacchetta:2000jk,Kanazawa:2013uia,Wei:2013csa,Wei:2014pma,Chen:2016moq}

Significant efforts have been devoted to extracting FFs of different hadron species from a variety of experimental measurements in the past decades. The unpolarized FFs have now been well constrained in these efforts~\cite{Binnewies:1994ju,deFlorian:1997zj,Kniehl:2000fe,Kneesch:2007ey,Kretzer:2000yf,Albino:2005me,deFlorian:2007ekg,Aidala:2010bn,deFlorian:2014xna,deFlorian:2017lwf}. However, the extraction of the spin-dependent FFs is still at its infancy~\cite{deFlorian:1997zj,Anselmino:2007fs,Kang:2015msa,Anselmino:2015sxa,Anselmino:2015fty,Chen:2016iey,Chen:2020pty}, which usually relies on colliders with polarized beams or high-energy reactions dominated by weak interactions, and requires rather strict accuracy of experiments. Recently, the transverse spin transfer ($D^{\perp}_{1\rm T}$) has been measured at BELLE~\cite{Belle:2018ttu}, which opens the possibility of studying the spin dependence of FFs using neither polarized nor weak-interaction-dominating collisions \cite{DAlesio:2020wjq,Callos:2020qtu,Chen:2021hdn,Gamberg:2021iat,DAlesio:2022brl,DAlesio:2023ozw}. Then, a natural question is whether the longitudinal spin transfer ($G_{1\rm L}$) can be constrained in a similar manner.

The longitudinal spin transfer, which is in charge of passing the longitudinal polarization from parent partons to daughter hadrons, was first numerically extracted by the DSV parameterization~\cite{deFlorian:1997zj} from experimental data in polarized collisions. Later, the correlation of dihadron polarization in unpolarized $e^+e^-$~\cite{Chen:1994ar,Zhang:2023ugf} and $pp$~\cite{Zhang:2023ugf} collisions has been proposed as a novel probe which paves the way for exploring the longitudinal spin transfer in unpolarized collisions. Meanwhile, the spin-spin correlation of two hard hadrons has also been proposed as a tool to study the Belle inequality and quantum entanglement recently~\cite{Gong:2021bcp,Vanek:2023oeo,Tu:2023few}. While this observable helps involve more available data in analyzing the spin transfer, this method suffers from the limited luminosity of current experiments and thus is quite challenging in reality. Therefore, it would be of great interest to extend these studies to relativistic heavy-ion collisions where the luminosity is much higher.

One of the remarkable milestones in relativistic heavy-ion physics is the discovery of the strongly coupled quark-gluon plasma (QGP) matter. Energetic partons produced from the initial hard collisions lose a portion of their energy inside the QGP before fragmenting into hadrons. This is known as jet quenching and is considered smoking gun evidence of the formation of the QGP~\cite{Wang:1992qdg,Qin:2015srf,Majumder:2010qh,Blaizot:2015lma,Cao:2020wlm}. Significant advances have been achieved over the past three decades in evaluating elastic and inelastic energy loss of jets inside the QGP~\cite{Bjorken:1982tu,Djordjevic:2006tw,Qin:2007rn,Baier:1994bd,Baier:1996kr,Zakharov:1996fv,Gyulassy:1999zd,Wiedemann:2000za,Arnold:2002ja,Wang:2001ifa}, and developing sophisticated Monte-Carlo event generators to understand the related phenomenology~\cite{Armesto:2009fj,Schenke:2009gb,Zapp:2013vla,Casalderrey-Solana:2014bpa,Cao:2017qpx,Luo:2023nsi,Putschke:2019yrg}. With these endeavors, studies on jets have now been extended from medium modification on jets to jet-induced medium excitation~\cite{Cao:2022odi,Bouras:2014rea,Casalderrey-Solana:2016jvj,Tachibana:2017syd,Chen:2017zte,Milhano:2017nzm}, and from understanding the jet-medium interaction mechanisms to utilizing jets to precisely extract properties of the QCD medium under extreme conditions~\cite{JET:2013cls,JETSCAPE:2021ehl,Feal:2019xfl,Liu:2023rfi,Karmakar:2023ity,Wu:2022vbu}. A recent study also shows that jets can acquire polarization while traversing an anisotropic medium~\cite{Hauksson:2023tze}. Nevertheless, discussion on the polarization of quenched jets is still rare in literature. The interplay between spin and jet quenching remains mostly uncharted territory. This will be the main focus of our present study. On the other hand, as a result of the global polarization effect~\cite{Liang:2004ph,Liang:2004xn,Gao:2007bc,Becattini:2016gvu,Florkowski:2018fap,STAR:2018gyt,Becattini:2020sww,
Becattini:2021suc}, the polarization of two soft hadrons formed in the QGP can be also correlated \cite{STAR:2017ckg,STAR:2019erd,Ivanov:2020wak,Ryu:2021lnx,STAR:2023nvo}. Therefore, it is important not to confuse the polarization correlation of two large-$p_{\rm T}$ hadrons produced through hard scatterings in this work with that of soft QGP hadrons. They reside in different kinematic regions and originate from different effects. 

Furthermore, another intriguing aspect of heavy-ion collisions is the ultraperipheral collisions (UPC), whose impact parameter is so large that nucleons from projectile and target beams can hardly scatter through the strong force. Instead, the two nuclei may interact via exchanging quasi-real photons or pomerons. This has attracted an increasing amount of interest over the past few years~\cite{Bertulani:1987tz,Baltz:2007kq,Contreras:2015dqa,STAR:2018ldd,Li:2019sin,Li:2019yzy,Hatta:2020bgy,Hagiwara:2021xkf,Zhou:2022twz,Shao:2022stc,Shao:2023zge,Iancu:2023lel}. Recently, ATLAS and CMS collaborations at the LHC \cite{ATLAS:2017kwa,CMS:2020ekd,ATLAS:2022cbd,CMS:2022lbi} demonstrated the capability of measuring back-to-back dijets in the UPC process with transverse momenta as high as $10\sim 20$ GeV. Several theoretical studies have been carried out~\cite{Hagiwara:2017fye,Mantysaari:2019csc,Hatta:2020bgy,Guzey:2020ehb,Shao:2023zge,Iancu:2023lel}. Although jet quenching is absent in UPC, this provides an alternative site for studying the spin-dependent FFs that is on a par with the $e^+e^-$ annihilation process but offers complementary kinematic regions. 

The rest of this paper will be organized as follows. In Sec.~\ref{section2}, we will first develop the formalism of studying the dihadron polarization that involves the parton energy loss inside the QGP, and then systematically explore the helicity correlation between hyperon pairs from $pp$ collisions to peripheral and central heavy-ion collisions. In Sec.~\ref{section3}, we will investigate this correlation in UPC where contributions from photon-photon and photon-pomeron interactions will be discussed separately. Section~\ref{section4} presents a summary in the end.
	
\section{Dihadron polarization in central and peripheral heavy-ion collisions}
\label{section2}

\subsection{Correlation of dihardron polarization in relativistic heavy-ion collisions}

We consider the following two processes that generate $\Lambda$ ($\bar{\Lambda}$) hyperons in AA collisions:
\begin{equation}
\begin{aligned}
\mathrm{A}+\mathrm{A} \rightarrow \Lambda(\lambda_1,\eta_1,\vec{p}_{\rm{T1}}) + {\Lambda} (\lambda_2,\eta_2,\vec{p}_{\rm{T2}}) +X, \\ 
\mathrm{A}+\mathrm{A} \rightarrow \Lambda(\lambda_1,\eta_1,\vec{p}_{\rm{T1}}) + \bar{\Lambda} (\lambda_2,\eta_2,\vec{p}_{\rm{T2}}) +X,
\end{aligned}
\end{equation}
in which a pair of almost back-to-back hard partons are first created from an energetic nucleon-nucleon collision and then fragment to $\Lambda$ (or $\bar{\Lambda}$) hyperons respectively. We use $\lambda_{1(2)}=\pm 1$, $\eta_{1(2)}$ and $\vec{p}_\mathrm{T1(2)}$ to denote the helicity, rapidity and transverse momentum of the final state hyperons. Following the earlier study on $pp$ collisions~\cite{Zhang:2023ugf}, we write the cross section of such dihadron production in AA collisions as~\cite{Wang:2016fds}
\begin{equation}
\label{eq:xsection}
\begin{aligned}
\frac{d\sigma^{\rm{AA}}_{\lambda_1\lambda_2}}{d\eta_1 d^2 \vec{p}_{\rm{T1}} d\eta_2 d^2 \vec{p}_{\rm{T2}} } & = \int dxdy T_{\rm{A}}(x-\tilde{b}/2,y)T_{\rm{A}}(x+\tilde{b}/2,y)  \int \frac{dz_1}{z^2_1} \frac{dz_2}{z^2_2} \sum_{ab \rightarrow cd} \sum_{\lambda_{c} \lambda_{d}} \frac{1}{\pi}   x_a f_{{\rm A},a}(x_a) x_bf_{{\rm A},b}(x_b) \\
& \times \frac{d \hat{\sigma} ^{ab \rightarrow cd} _{\lambda_{c} \lambda_{d}} }{dt} \,\mathcal{D}_{c}(z_1,\lambda_1;\lambda_{c}) \mathcal{D}_{d} (z_2, \lambda_2; \lambda_{d}) 
\,\delta^2\left(\frac{\vec{p}_{\rm{T1}} }{z_1} + \Delta \vec{p}_{\rm{T1}} + \frac{\vec{p}_{\rm{T2}} }{z_2} + \Delta \vec{p}_{\rm{T2}}\right) +  \big\{c \leftrightarrow d  \big\}.
\end{aligned}
\end{equation}
Here, we place the centers of the two colliding nuclei at $(\pm{\tilde b}/2,0,0)$ with the nuclear thickness function given by
\begin{equation}
T_{\rm{A}}(x,y) = \int dz \frac{\rho_0}{1+e^{\left(\sqrt{x^2+y^2+z^2}-R\right)/a}},
\end{equation}
where $\rho_0 = 0.17$~fm$^{-3}$, $R_{\rm{Au}} = 6.38$~fm, $a_{\rm{Au}} = 0.535$~fm, $R_{\rm{Pb}} = 6.62$~fm and $a_{\rm{Pb}} = 0.546$~fm are the parameters for the Woods-Saxon distribution of nucleon density inside Au and Pb nuclei. In Eq.~(\ref{eq:xsection}), $f_{{\rm A},a(b)}$ represents the parton distribution function inside a nucleus (nPDF), which is obtained by convoluting the collinear CTEQ PDF~\cite{Dulat:2015mca} with the EPPS parameterization of the cold nuclear matter effect~\cite{Eskola:2016oht}; $d \hat{\sigma}^{ab \rightarrow cd} _{\lambda_{c} \lambda_{d}} / dt $ is the differential partonic hard cross section with respect to the Mandelstam variable $t$; and $\mathcal{D}_{c(d)}$ denotes the helicity-dependent FF. {At the leading order, we assume the two final hyperons are antiparallel to each other in the transverse plane. This yields the $\delta$-function that requires the back-to-back production of partons $c$ and $d$: $\vec{p}_{\rm{T1}} /z_1 + \Delta \vec{p}_{\rm{T1}} = -( \vec{p}_{\rm{T2}} / z_2 + \Delta \vec{p}_{\rm{T2}})$, where $\Delta \vec{p}_{\rm{T}1(2)}$ is the energy (or momentum) loss of a hard parton inside the QGP before fragmenting into a hadron with a momentum fraction $z_{1(2)}$. The momentum loss here depends on the production location $(x,y)$ of the hard parton, as well as its initial energy and its path through the QGP. This introduces the main difference between our present study on AA collisions and the prior work on $pp$ collisions~\cite{Zhang:2023ugf}, and will be discussed in detail in the next subsection. We ignore the transverse momentum broadening of hard partons inside the QGP and during fragmentation, and thus their initial momentum, momentum loss and the final hadron momentum are parallel to each other. The fractional momentum taken by partons $a$ and $b$ from their parent nucleons are then restricted by $x_a={(p_{\rm{T1}}/z_1 + \Delta {p}_{\rm{T1}})} (e^{\eta_1} + e^{\eta_2})/{\sqrt{s_\mathrm{NN}}}$ and $x_b={(p_{\rm{T1}}/z_1 + \Delta {p}_{\rm{T1}})}  (e^{-\eta_1} + e^{-\eta_2})/{\sqrt{s_\mathrm{NN}}}$, where $\sqrt{s_\mathrm{NN}}$ is the center-of-mass energy per nucleon pair.} Since we cannot determine which parton fragments to the first hyperon, the exchange between $\{c \leftrightarrow d\}$ is needed for non-identical final state partons. In the end, the dihadron cross section is obtained by summing over all possible hard scattering channels $ab \rightarrow cd$ and the intermediate parton helicities $\lambda_c$ and $\lambda_d$. Eight types of partonic channels are taken into account: $q_i + q_j \rightarrow q_i + q_j$, $q_i + q_i \rightarrow q_i + q_i$, $q_i + \bar{q}_i \rightarrow q_i + \bar{q}_i$, $q_i + \bar{q}_i \rightarrow q_j + \bar{q}_j$, $g+g \rightarrow q_i + \bar{q}_i$, $q_i + \bar{q}_i \rightarrow g+g$, $q_i + g \rightarrow q_i +g$ and $g+g \rightarrow g+g$, where $q$ ($\bar{q}$) represents quarks (anti-quarks) with subscripts ($i$ or $j$) denoting their flavors, and $g$ represents gluons. {Their helicity-dependent differential cross sections are listed in Ref.~\cite{Zhang:2023ugf}, with the Mandelstam variables of these scatterings given by $s=x_ax_b s_{\rm{NN}}$, $t=-x_a \sqrt{s_{\rm{NN}}} ({p_{\rm{T1}}}/{z_1}+\Delta p_{\mathrm{T}1}) e^{- \eta_1}$, $u=-x_a \sqrt{s_{\rm{NN}}} ({p_{\rm{T1}}}/{z_1}+\Delta p_{\mathrm{T}1}) e^{- \eta_2}$.} We consider four flavors of quarks -- $u$, $d$, $s$, $c$ -- in this work.

In the current analysis, we only focus on the energy loss aspect of the medium effect. In principle, the medium could also directly alter the polarization of a high energy parton. However, this effect has a minor impact on our study. This conclusion can be obtained by comparing the polarized splitting functions derived in Ref.~\cite{Ravindran:1996ri} and their unpolarized partners. While the quark will definitely not lose any of its polarization due to the helicity conservation, the high-energy gluon will also not lose much polarization as long as it does not lose a significant portion of its energy while traversing the hot medium. Furthermore, the global orbital angular momentum deposited into the QGP in non-central heavy-ion collisions and high energy jet partons can also induce polarization of low energy medium partons, which will eventually hadronize into soft hadrons. These are beyond the scope of our current study on energetic hadrons from jet fragmentation.

To study the longitudinal spin transfer from parton to hadron, one may decompose the FFs as~\cite{Zhang:2023ugf}
 \begin{equation}
 \label{eq:Ddecomp}
 \begin{aligned}
& \mathcal{D}_{c}(z_1,\lambda_1;\lambda_c) = D_{1,c}(z_1) + \lambda_1 \lambda_c G_{1{\rm L},c}(z_1),\\
& \mathcal{D}_{d}(z_2,\lambda_2;\lambda_d) = D_{1,d}(z_2) + \lambda_2 \lambda_d G_{1{\rm L},d}(z_2),
 \end{aligned}
 \end{equation}
where $D_{1,c(d)}$ is the unpolarized FF and $G_{1{\rm L},c(d)}$ is the longitudinal spin transfer at the leading twist. By substituting Eq.~(\ref{eq:Ddecomp}) into Eq.~(\ref{eq:xsection}) and integrating over the transverse momenta of the two hyperons, we obtain
\begin{equation}
\label{eq:xsection2}
\begin{aligned}
\frac{d\sigma^{\rm{AA}}_{\lambda_1\lambda_2}}{d\eta_1 d\eta_2} & = \int dxdy T_{\rm{A}}(x-\tilde{b}/2,y)T_{\rm{A}}(x+\tilde{b}/2,y)  \int dz_1 dz_2 \frac{1}{z^2_1} \int d^2 \vec{p}_{\rm{T1}} \sum_{ab \rightarrow cd}  \frac{1}{\pi}   x_a f_{{\rm A},a}(x_a) x_bf_{{\rm A},b}(x_b) \\
& \times \left\{ D_{1,c}(z_1) D_{1,d}(z_2) \left[\frac{d\hat{\sigma}_{++}^{ab \rightarrow cd} }{dt} + \frac{d\hat{\sigma}_{--}^{ab \rightarrow cd} }{dt} + \frac{d\hat{\sigma}_{+-}^{ab \rightarrow cd} }{dt} + \frac{d\hat{\sigma}_{-+}^{ab \rightarrow cd} }{dt} \right]  \right.\\ 
&\left. +\, \lambda_1 \lambda_2 G_{{1\rm L},c}(z_1) G_{1{\rm L},d}(z_2) \left[ \frac{d\hat{\sigma}_{++}^{ab \rightarrow cd}}{dt} + \frac{d\hat{\sigma}_{--}^{ab \rightarrow cd}}{dt} - \frac{d\hat{\sigma}_{+-}^{ab \rightarrow cd}}{dt} - \frac{d\hat{\sigma}_{-+}^{ab \rightarrow cd}}{dt} \right] \right\} +  \big\{c \leftrightarrow d  \big\}.
\end{aligned}
\end{equation}
Here, the $\delta$-function in Eq.~(\ref{eq:xsection}) has been used to complete the integral over $d\vec{p}_{{\rm T}2}$, and the parton energy loss inside the QGP is implicitly included in the calculation of the $x_{a(b)}$ variables and the partonic cross sections. The cross terms $D_{1,c}(z_1)G_{1{\rm L},d}(z_2)$ and $D_{1,d}(z_2)G_{{1\rm L},c}(z_1)$ vanish since $d\hat{\sigma}_{++}^{ab \rightarrow cd}/{dt}=d\hat{\sigma}_{--}^{ab \rightarrow cd}/{dt}$ and $d\hat{\sigma}_{+-}^{ab \rightarrow cd}/{dt}=d\hat{\sigma}_{-+}^{ab \rightarrow cd}/{dt}$ in the partonic cross sections~\cite{Zhang:2023ugf}. 

With Eq.~(\ref{eq:xsection2}) above, we can define the correlation function of dihadron polarization as the probability of the two hadrons taking the same helicity minus that taking opposite helicities: 
\begin{align}
\label{AAC}
\mathcal{C}_{\rm{LL}}&(\eta_1,\eta_2) = \left. \left( \frac{d\sigma^{\rm{AA}}_{\lambda_1=\lambda_2}}{d\eta_1 d\eta_2} - \frac{d\sigma^{\rm{AA}}_{\lambda_1=-\lambda_2}}{d\eta_1 d\eta_2} \right) \middle/ \left( \frac{d\sigma^{\rm{AA}}_{\lambda_1=\lambda_2}}{d\eta_1 d\eta_2} + \frac{d\sigma^{\rm{AA}}_{\lambda_1=-\lambda_2}}{d\eta_1 d\eta_2} \right) \right.\\
= &\, \frac{ \bigintsss dx dy dz_1 dz_2 d^2 \vec{p}_{\rm{T1}} T_{\rm{A}}\left(x-\frac{\tilde{b}}{2},y\right)T_{\rm{A}}\left(x+\frac{\tilde{b}}{2},y\right) \sum\limits_{ab \rightarrow cd} \dfrac{x_a x_b}{z_1^2\pi}  f_{{\rm A},a}(x_a)  f_{{\rm A},b}(x_b) \dfrac{d\hat{\sigma}^{\rm{dif}}}{dt} G_{{1\rm L},c}(z_1) G_{{1\rm L},d}(z_2) + \big\{c \leftrightarrow d \big\} } { \bigintsss dx dy dz_1 dz_2 d^2 \vec{p}_{\rm{T1}} T_{\rm{A}}\left(x-\frac{\tilde{b}}{2},y\right)T_{\rm{A}}\left(x+\frac{\tilde{b}}{2},y\right) \sum\limits_{ab \rightarrow cd} \dfrac{x_a x_b}{z_1^2\pi}  f_{{\rm A},a}(x_a)  f_{{\rm A},b}(x_b) \dfrac{d\hat{\sigma}^{\rm{sum}}}{dt} D_{{1},c}(z_1) D_{{1},d}(z_2) + \big\{c \leftrightarrow d \big\} }, \nonumber
\end{align}
with
\begin{align}
\frac{d \hat{\sigma} ^{\rm{dif}}}{dt}= \frac{d\hat{\sigma}_{++}^{ab \rightarrow cd}}{dt} + \frac{d\hat{\sigma}_{--}^{ab \rightarrow cd}}{dt} - \frac{d\hat{\sigma}_{+-}^{ab \rightarrow cd}}{dt} - \frac{d\hat{\sigma}_{-+}^{ab \rightarrow cd}}{dt},\\
\frac{d \hat{\sigma} ^{\rm{sum}}}{dt}= \frac{d\hat{\sigma}_{++}^{ab \rightarrow cd}}{dt} + \frac{d\hat{\sigma}_{--}^{ab \rightarrow cd}}{dt} + \frac{d\hat{\sigma}_{+-}^{ab \rightarrow cd}}{dt} + \frac{d\hat{\sigma}_{-+}^{ab \rightarrow cd}}{dt}.
\end{align}
This is the main quantity that we will examine in the rest of this work.
	
\subsection{Parton energy loss inside the QGP}
	
As discussed in previous subsection, in relativistic heavy-ion collisions, hard partons suffer energy loss inside the QGP before fragmenting into hadrons. This introduces the main difference in the dihadron correlation between AA and $pp$ collisions. In this work, we use a linear Boltzmann transport (LBT) model~\cite{Cao:2016gvr,Luo:2023nsi} to calculate the parton energy loss in high-energy nuclear collisions.

In the LBT model, the evolution of the phase space distribution of energetic partons (denoted by ``$a$") -- $f_a (\tilde t, \vec{x}_a, \vec{p}_a)$ -- is described by the following Boltzmann equation 
\begin{equation}
p_a \cdot \partial f_a(\tilde t, \vec{x}_a,\vec{p}_a)= E_a (C_\mathrm{el}[f_a]+C_\mathrm{inel}[f_a]),
\label{eq:Boltzmann}
\end{equation}
where $p_a=(E_a, \vec{p}_a)$ denotes the four momentum and $\vec x_a$ denotes the position. To avoid conflict with the Mandelstam variable $t$, we use $\tilde t$ to represent the time in this paper. Both elastic and inelastic scatterings between $a$ and the QGP medium contribute to the collision integral, $C_\mathrm{el}[f_a]$ and $C_\mathrm{inel}[f_a]$, on the right hand side. If one only considers the medium modification on the parton $a$ but neglects the inverse process, the expression above can be viewed as a linear equation of $f_a$. 

For an elastic scattering $ab\rightarrow cd$, where $b$ represents a medium parton, $c$ and $d$ are the final states of $a$ and $b$ respectively, the scattering rate can be extracted from the collision integral as
\begin{align}
\label{eq:rate}
\Gamma_a^\mathrm{el}(E_a,T)=&\sum_{b,(cd)}\frac{\gamma_b}{2E_a}\int \prod_{i=b,c,d}\frac{d^3p_i}{E_i(2\pi)^3} f_b(E_b,T) [1\pm f_c(E_c,T)][1\pm f_d(E_d,T)] S_2(s,t,u)\nonumber\\
&\times (2\pi)^4\delta^{(4)}(p_a+p_b-p_c-p_d)|\mathcal{M}_{ab\rightarrow cd}|^2,
\end{align}
in which $T$ is the local temperature of the medium, $\gamma_b$ is the color-spin degrees of freedom of parton $b$, and $f_{b,c,d}$ takes the Bose (Fermi) distribution of gluons (quarks) inside the QGP, with the Bose enhancement (Fermi suppression) effect on the final state taken into account by the $\pm$ notation. The function $S_2(s,t,u)=\theta(s\ge 2\mu_\mathrm{D}^2)\,\theta(-s+\mu^2_\mathrm{D}\le t \le -\mu_\mathrm{D}^2)$ is introduced to avoid the collinear divergence in the leading-order (LO) scattering matrices $\mathcal{M}_{ab\rightarrow cd}$, where $s$, $t$ and $u$ are the Mandelstam variables and $\mu_{\rm{D}}^2=g^2T^2(N_c + N_f/2)/3$ is the Debye screening mass with $g^2=4\pi \alpha_\mathrm{s}$ being the strong coupling constant, $N_c=3$ and $N_f=3$ being the color and flavor numbers respectively. Here, we sum over contributions from all possible scattering channels, whose matrix elements have been summarized in Ref.~\cite{Auvinen:2009qm}. 

For inelastic scattering, we relate its rate to the average number of medium-induced gluons per unit time as
\begin{equation}
\label{eq:gluonnumber}
\Gamma_a^\mathrm{inel} (E_a,T,\tilde{t}\,) = \int dzdk_\perp^2 \frac{1}{1+\delta^{ag}}\frac{dN_g^a}{dz dk_\perp^2 d\tilde{t}},
\end{equation}
where the emitted gluon spectrum is adopted from the higher-twist energy loss calculation~\cite{Wang:2001ifa,Zhang:2003wk,Majumder:2009ge},
\begin{equation}
\label{eq:gluondistribution}
\frac{dN_g^a}{dz dk_\perp^2 d\tilde{t}}=\frac{2\alpha_\mathrm{s} C_A P_a(z)}{\pi k_\perp^4}\,\hat{q}_a \left(\frac{k^2_{\perp} }{k^2_{\perp} + z^2m_a^2 } \right)^4 {\sin}^2\left(\frac{\tilde{t}-\tilde{t}_i}{2\tau_f}\right).
\end{equation}
In the equations above, $z$ and $k_\perp$ are the fractional energy and transverse momentum of the emitted gluon with respect to its parent parton. We use $\tilde{t}_i$ to represent the production time (or the time of the previous splitting) of parton $a$. In addition, $C_A=N_c=3$ is the color factor, $P_a(z)$ and $m_a$ are the splitting function and mass of parton $a$ respectively, and $\tau_f=2E_a z(1-z)/(k_\perp^2+z^2 m_a^2)$ is the formation time of the emitted gluon. We take zero mass for gluons and light flavor quarks, and the bare mass $1.27$~GeV for charm quarks in this work. The medium information is embedded in the jet transport parameter $\hat{q}_a$, defined as the transverse momentum broadening square per unit time of parton $a$ due to elastic scatterings -- $\hat{q}_a=d\langle q_\perp^2\rangle_a/d\tilde{t}$, which can be evaluated via Eq.~(\ref{eq:rate}) with a weight factor of $q_\perp^2=[\vec{p}_c-(\vec{p}_c\cdot \hat{\vec p}_a)\hat{\vec p}_a]^2$ added to its right. Here, $\hat{\vec p}_a\equiv = \vec p_a / |\vec p_a|$. In Eq.~(\ref{eq:gluonnumber}), lower and higher cutoffs $z_\mathrm{min}=\mu_\mathrm{D}/E_a$ and $z_\mathrm{min}=1-\mu_\mathrm{D}/E_a$ are implemented to avoid possible divergence in the integral. For the $g\rightarrow gg$ process, a 1/2 factor is introduced by the Kronecker delta function to prevent double counting in the splitting rate.

\begin{figure}[tbp!]
	\begin{center}
		\includegraphics[width=0.4\textwidth]{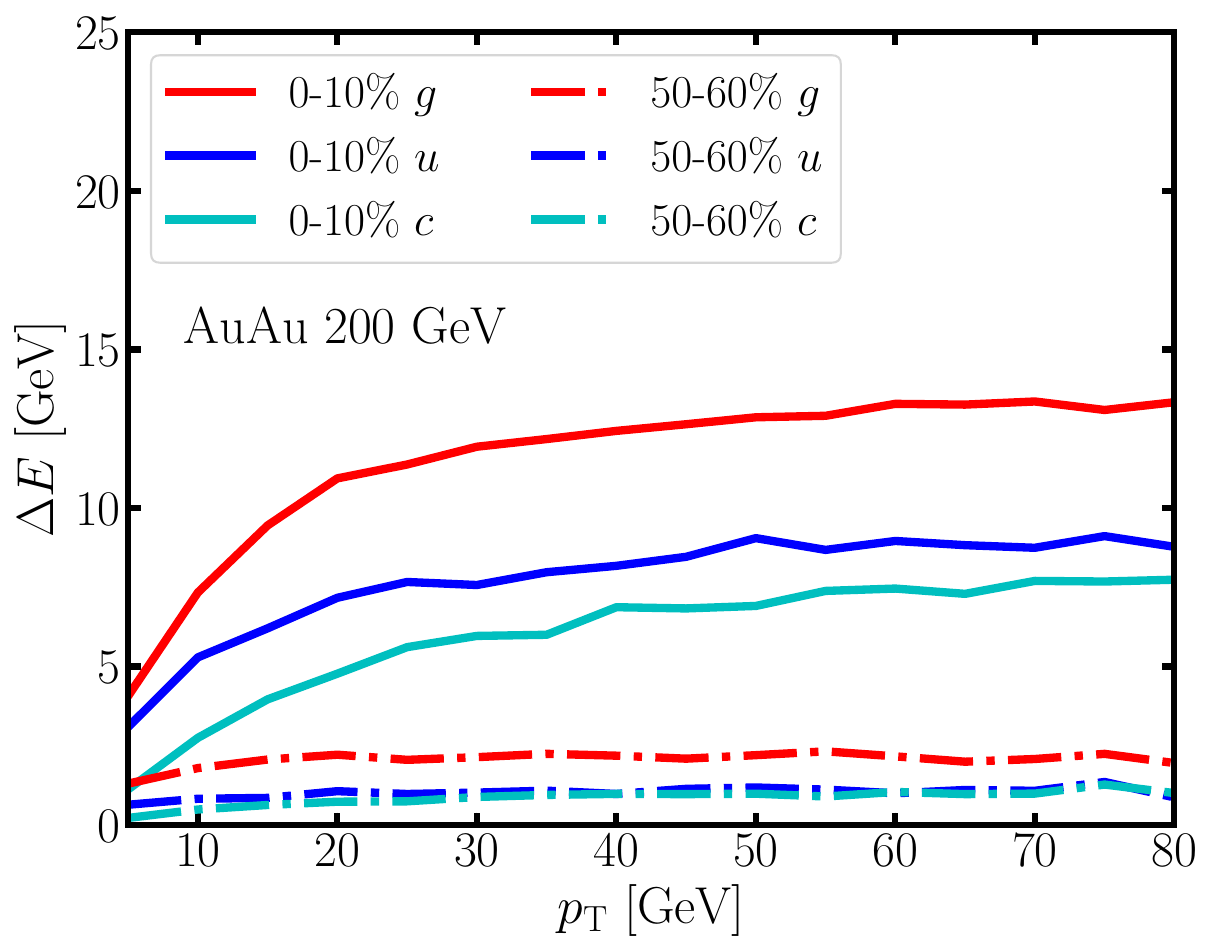}
		\includegraphics[width=0.4\textwidth]{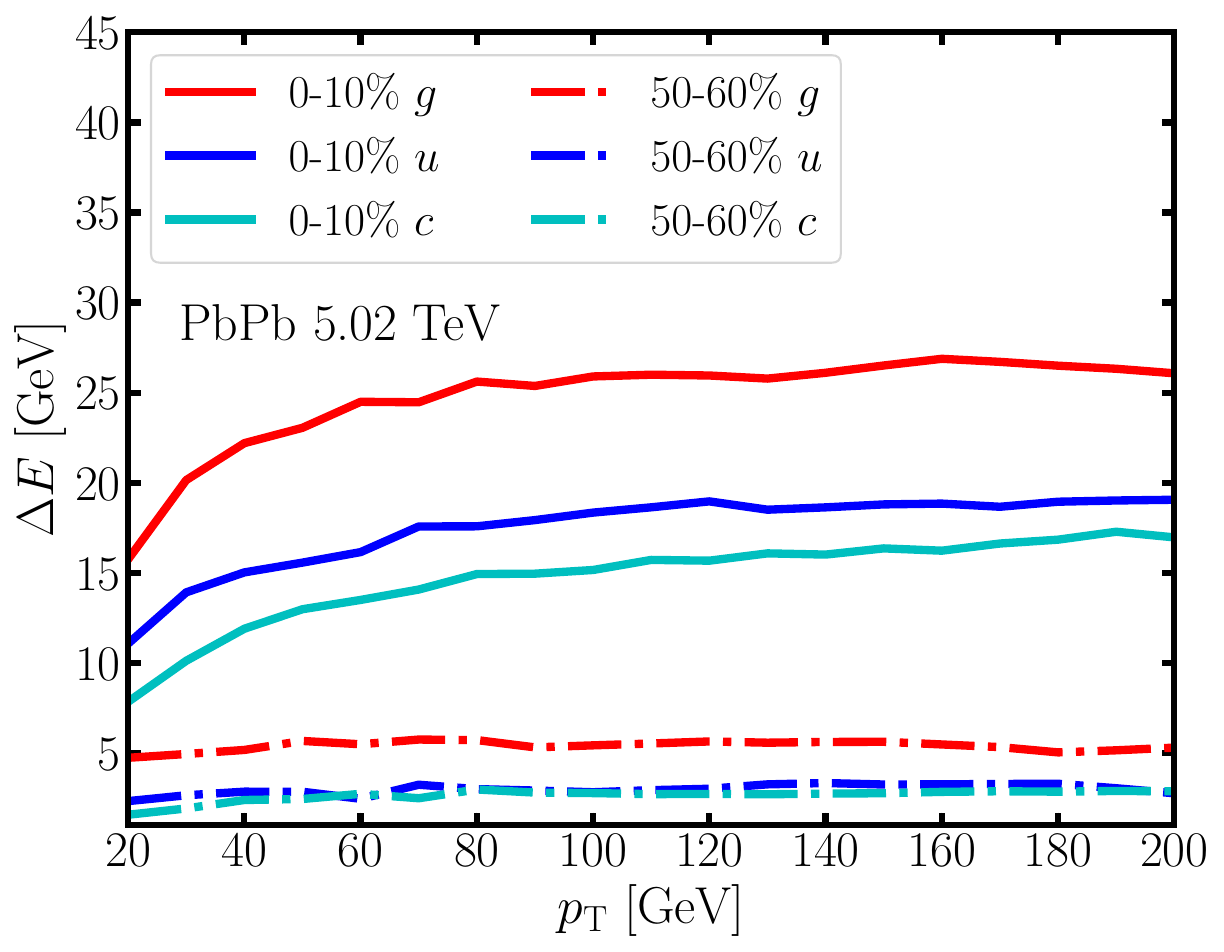}			
	\end{center}	
	\caption{(Color online) Average energy loss of different flavors of partons through the QGP at RHIC (left) and LHC (right), as a function of their initial $p_\mathrm{T}$. The partons are initialized at the center of the medium ($x=y=0$), and travel along the $+\hat{x}$ direction.}
	\label{dE}
\end{figure}

Using these scattering rates, one may conduct Monte-Carlo simulation of hard parton interactions with a thermal medium. For realistic heavy-ion collisions, we use a (3+1)-dimensional viscous hydrodynamic model CLVisc~\cite{Pang:2018zzo,Wu:2018cpc,Wu:2021fjf} to generate the spacetime evolution profiles of the QGP. At a given time step, we use the velocity field information of the hydrodynamic medium to boost each hard parton into the local rest frame of the QGP, in which we use the local temperature information to calculate the elastic and inelastic scattering rates between the parton and medium. After simulating the scattering processes according to their integrated and differential rates, we boost the final state partons back into the global frame where these partons stream freely to their locations at the next time step. We initiate the parton-medium interaction at the starting time of the hydrodynamic evolution (0.6~fm/$c$), and iterate the algorithm above until parton exits the medium (with local temperature below 165~MeV). In this work, we apply the model parameters of LBT used in Ref.~\cite{Xing:2019xae}, which provide a simultaneous description of nuclear modification factors of light and heavy flavor hadrons across a wide $p_\mathrm{T}$ region in heavy-ion collisions.
	
{Given the initial position ($x,y$) and momentum ($\vec{p}_\mathrm{T}$) of a hard parton in the transverse plane (with rapidity set as 0), we use the LBT model to simulate the scatterings of this parton through the QGP based on the rates given by Eqs.~(\ref{eq:rate}) and~(\ref{eq:gluonnumber}). Details on the model implementation can be referred to Ref.~\cite{Luo:2023nsi}. The energy loss ($\Delta E$) of this parton is then calculated as the energy difference between its initial and final states (before and after traversing the QGP). For a parton with a given initial state, its $\Delta E$ is obtained by averaging over thousands of events through the LBT simulation.} Shown in Fig.~\ref{dE} is the average energy loss of different flavors of partons in Au+Au collisions at $\sqrt{s_\mathrm{NN}}=200$~GeV (left panel) and Pb+Pb collisions at $\sqrt{s_\mathrm{NN}}=5.02$~TeV. Here, the partons are initialized at the center of the medium ($x=y=0$) with their transverse momentum ($p_\mathrm{T}$) along the $+\hat{x}$ direction.  From the figure, one can observe the parton energy loss first increases with the parton $p_\mathrm{T}$ and then saturates at high $p_\mathrm{T}$. These hard partons lose more energy in central (0-10\%) than in peripheral (50-60\%) collisions. Within the same centrality class, they experience stronger energy loss in more energetic collisions. Comparing between different parton species, we see gluons lose more energy than light flavor quarks due to the larger color factor of the former, and light flavor quarks lose more energy than charm quarks due to the mass effect in parton energy loss. This function of average parton energy loss, in terms of $x$, $y$ and the initial parton $\vec{p}_\mathrm{T}$, is then applied in Eq.~(\ref{AAC}) to obtain the correlation function of dihadron polarization in central to peripheral AA collisions. {Since we focus on results at mid-rapidity in this work, $\Delta E$ extracted here is used to approximate the magnitude of $\Delta {p}_\mathrm{T}$ in Eq.~(\ref{eq:xsection}).}

\subsection{Numerical results}

There are three scenarios of DSV parameterizations of the FFs to $\Lambda$ ($\bar{\Lambda}$) hyperons~\cite{deFlorian:1997zj}, which provide the same unpolarized parts of the FFs but different polarized parts. In Scenario 1, only $s$ ($\bar{s}$) quarks are responsible for the longitudinal spin transfer to $\Lambda$ ($\bar{\Lambda}$); while in Scenario 2, apart from the positive contribution from $s$ quarks, $u$ and $d$ quarks have small but negative contributions to this spin transfer; and in Scenario 3, $u$, $d$ and $s$ quarks share the same polarized parts of the FFs. While gluons and $c$ quarks directly contribute to the unpolarized parts of FFs, they are neglected in the polarized parts at the initial factorization scale and can accumulate contributions through DGLAP evolution. The contribution from gluons to the longitudinal spin transfer could be important at large scale, though the contribution from $c$ quarks remains negligible in this work.

\begin{figure}[tbp!]
	\begin{center}
		\includegraphics[width=0.4\textwidth]{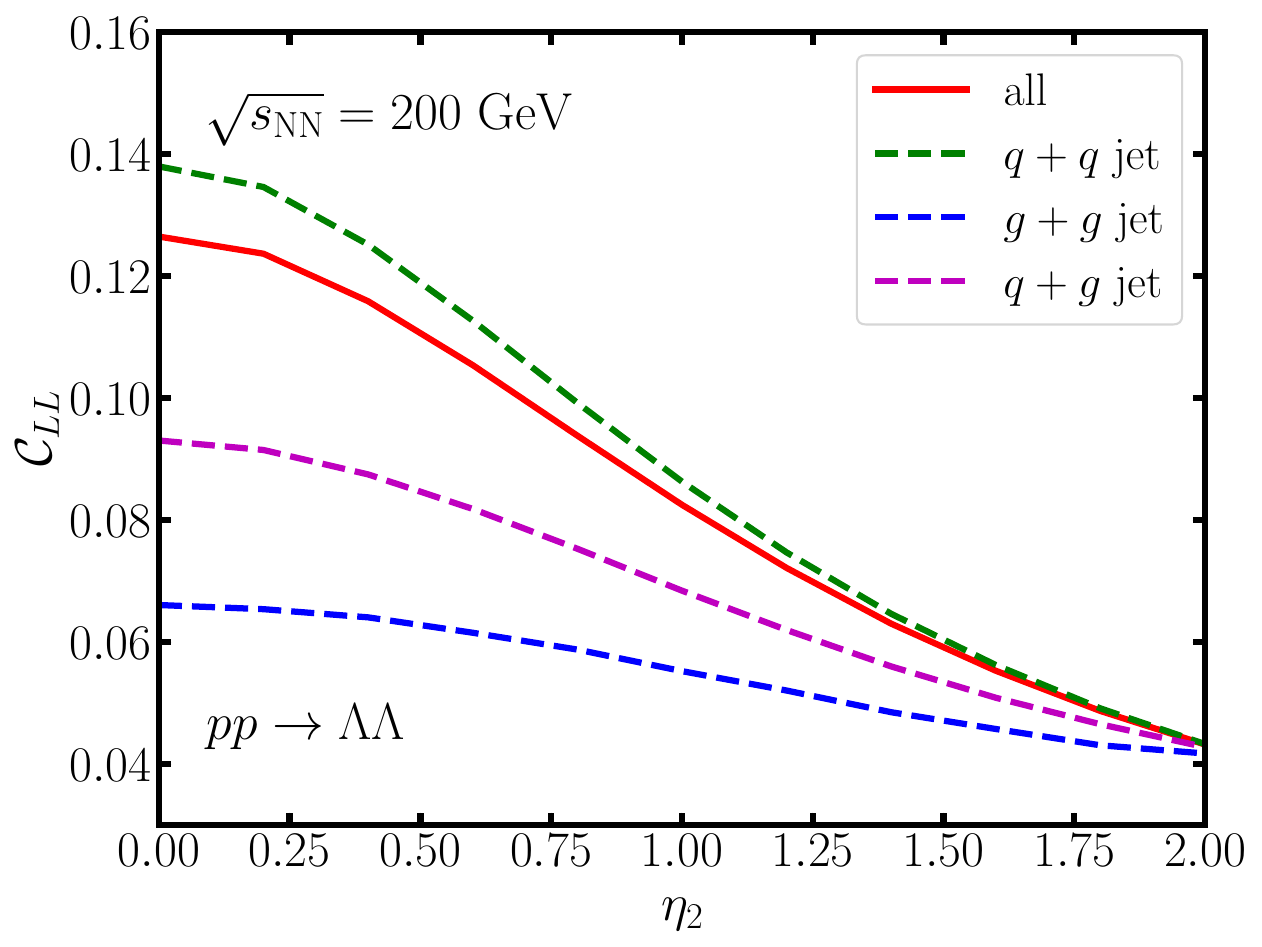}		
		\includegraphics[width=0.4\textwidth]{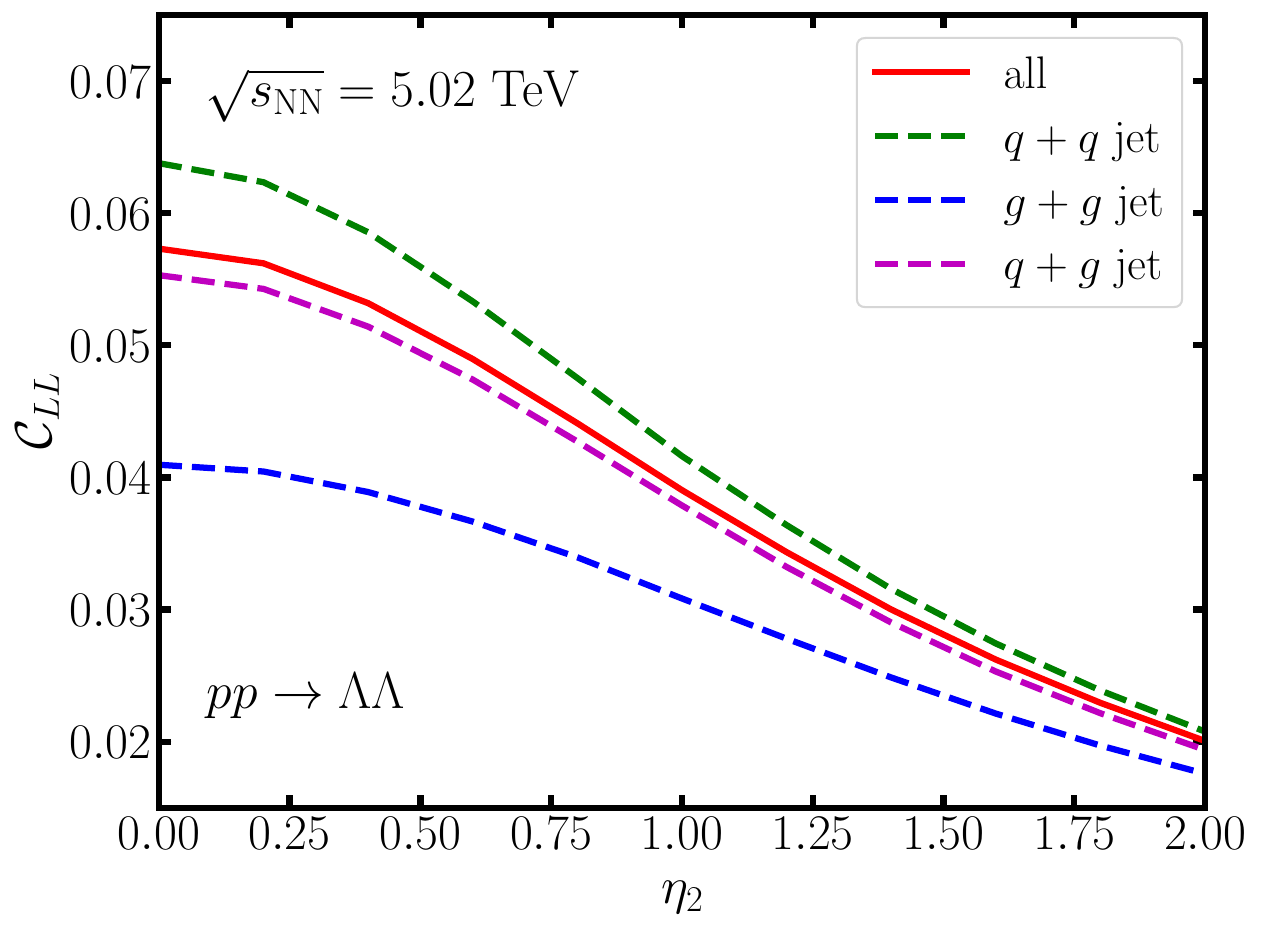}	
	\end{center}	
	\caption{(Color online) Correlation function of $\Lambda$-$\Lambda$ polarization in $pp$ collisions at RHIC (left) and LHC (right) energies, compared between results from different hard scattering channels and their combination.}
	\label{ppjet}
\end{figure}

We start with reviewing the dihadron polarization in $pp$ collisions in Fig.~\ref{ppjet} at RHIC (left panel) and LHC (right panel) energies, where the correlation functions between two $\Lambda$ hyperons are presented and compared between different categories of production channels. Here, we require one $\Lambda$ at mid-rapidity and plot $\mathcal{C}_\mathrm{LL}$ as a function of the other $\Lambda$'s rapidity. The transverse momenta of the $\Lambda$'s are restricted by $p_{\mathrm{T}1,2}>5$~GeV at RHIC and $p_{\mathrm{T}1,2}>20$~GeV at LHC. The DSV FFs with Scenario 3 is applied; and comparisons between different parameterizations of FFs will be shown later. Since the $\Lambda$-$\Lambda$ production is dominated by the $t$-channel process in nuclear collisions, which generally prefers same sign of the final parton helicities in the partonic cross sections, assuming FFs with Scenario 3 gives positive values of this correlation function~\cite{Zhang:2023ugf}. From the figure, one can observe two $\Lambda$'s from a pair of quark-quark jet possess the largest correlation of helicity, while that from a pair of gluon-gluon jet possess the smallest correlation. This is due to the smaller longitudinal spin transfer (or $G_{1\rm L}/D_1$) from gluons than from quarks in our calculation. We have verified that if the FFs are not convoluted in Eq.~(\ref{AAC}), one would obtain the inverse order, i.e., $\mathcal{C}_\mathrm{LL}^{q+q}<\mathcal{C}_\mathrm{LL}^{q+g}<\mathcal{C}_\mathrm{LL}^{g+g}$, due to different spin dependences of their partonic cross sections. Furthermore, we see larger values of $\mathcal{C}_\mathrm{LL}$ at RHIC than at LHC. This can be understood with the larger fractional momentum ($z$) region of the FFs and smaller contribution from the gluon jet probed at RHIC. Both effects result in a larger $G_{1\rm L}/D_{1}$ ratio.

\begin{figure}[tbp!]
	\begin{center}
		\includegraphics[width=0.4\textwidth]{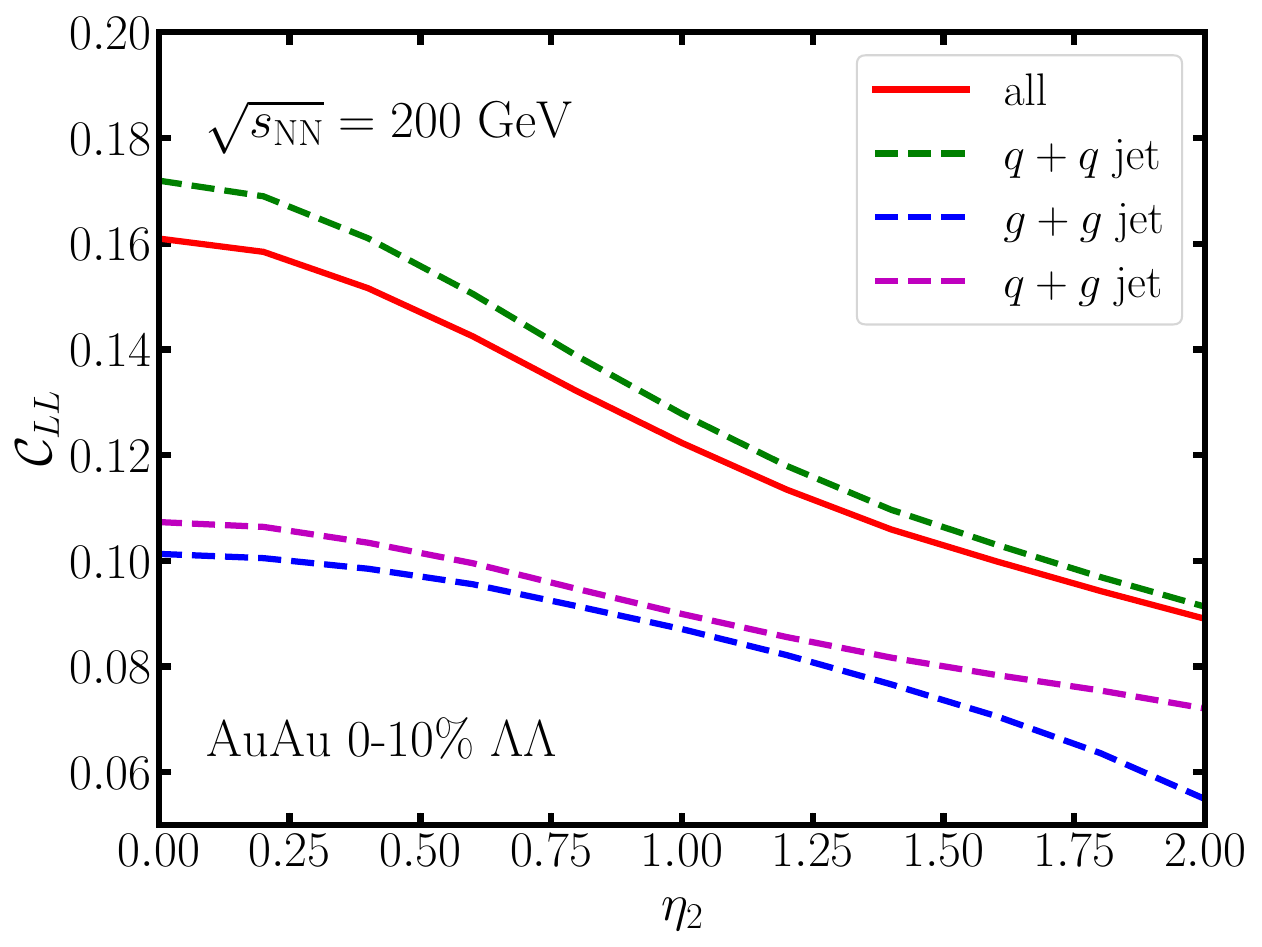}		
		\includegraphics[width=0.4\textwidth]{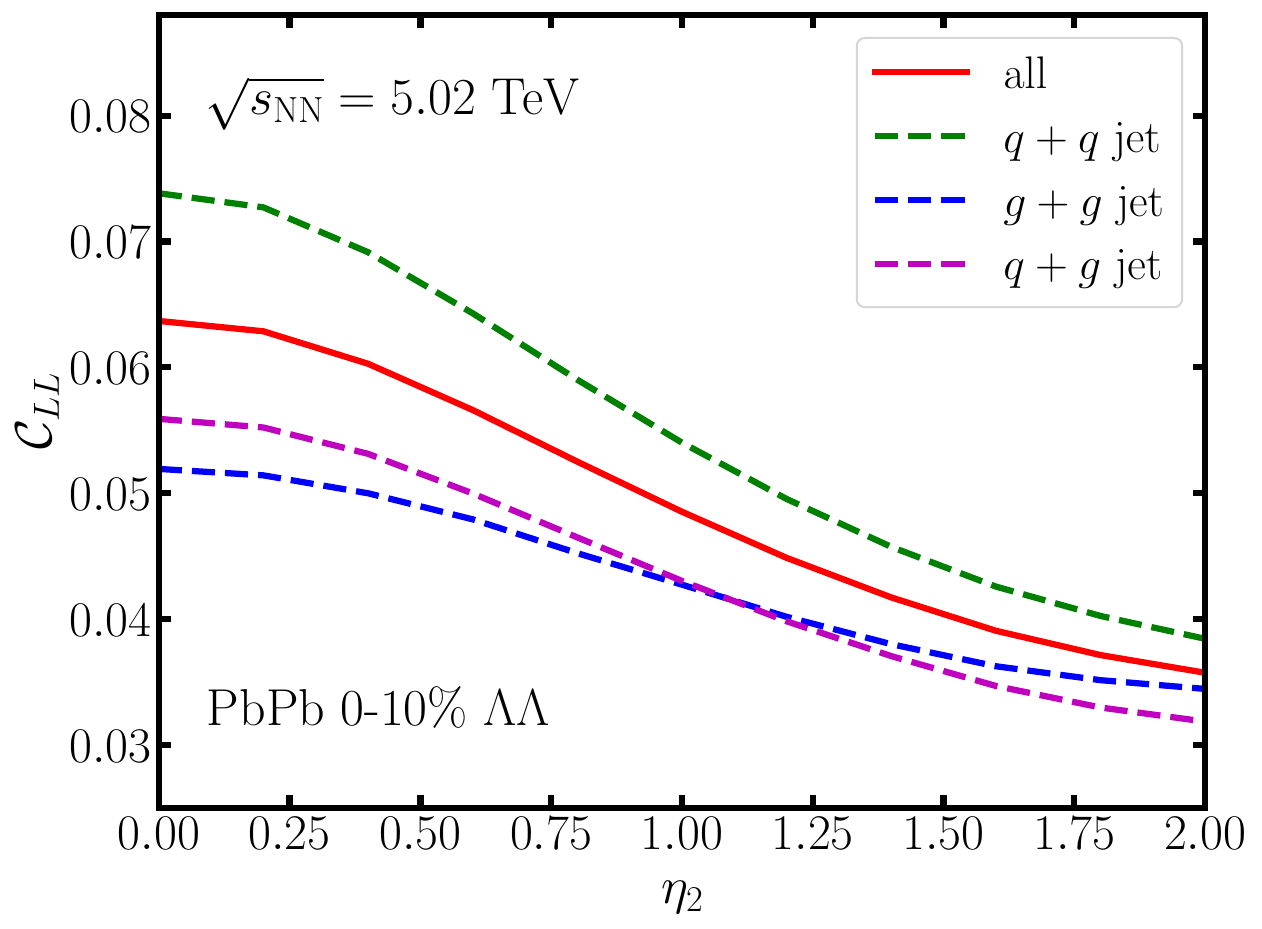}	
	\end{center}	
	\caption{(Color online) Correlation function of $\Lambda$-$\Lambda$ polarization in central Au+Au collisions at RHIC (left) and central Pb+Pb collisions at LHC (right), compared between results from different hard scattering channels and their combination.}
	\label{AAjet}
\end{figure}

In Fig.~\ref{AAjet}, we introduce the parton energy loss inside the QGP and present the correlation function of $\Lambda$-$\Lambda$ production in central Au+Au collisions at RHIC (left panel) and central Pb+Pb collisions at LHC (right panel). The kinematic regions of $\Lambda$'s are set in the same way as for the $pp$ collisions in Fig.~\ref{ppjet}. Due to parton energy loss, final $\Lambda$'s with the same $p_\mathrm{T}$ hadronize from partons with a lower average $p_\mathrm{T}$ in AA than in $pp$ collisions. This corresponds to a larger average $z$ in the FFs, and therefore a stronger longitudinal spin transfer in AA than in $pp$ collisions. Indeed, one can observe larger values of $\mathcal{C}_\mathrm{LL}$ in Fig.~\ref{AAjet} than in Fig.~\ref{ppjet} from each category of hard scatterings. This increment in $\mathcal{C}_\mathrm{LL}$ from $pp$ to AA collisions is more prominent for gluon-gluon jet than for quark-quark, considering the much stronger energy loss of gluons than quarks. Additionally, when the kinematics of the final state is fixed, energy loss increases the quark {\it vs.} gluon fractions produced by the initial hard collisions. This could be another reason why $\mathcal{C}_\mathrm{LL}$ is larger in AA than in $pp$ collisions after contributions from all channels are combined.

\begin{figure}[tbp!]
	\begin{center}
		\includegraphics[width=0.4\textwidth]{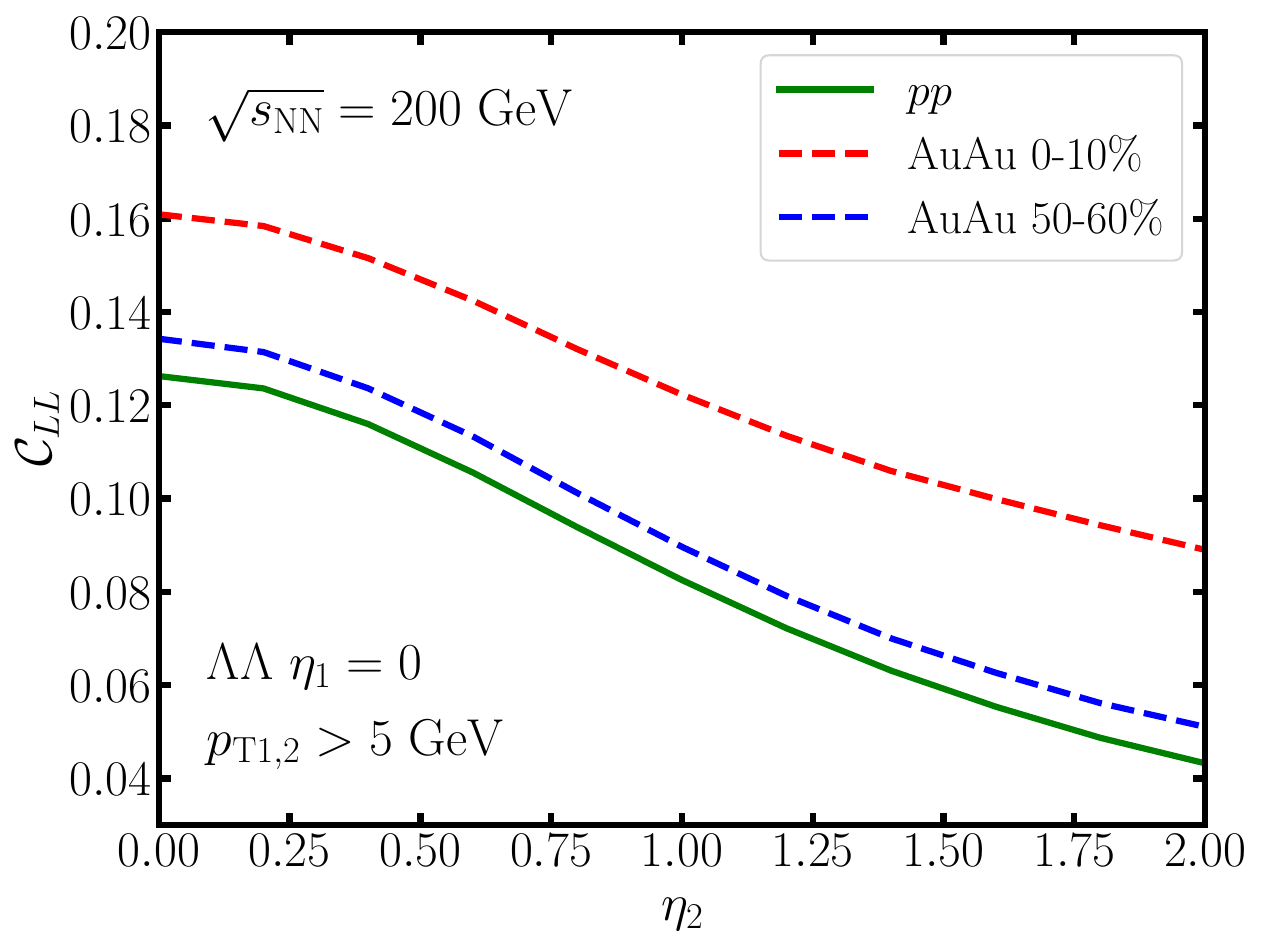}		
		\includegraphics[width=0.4\textwidth]{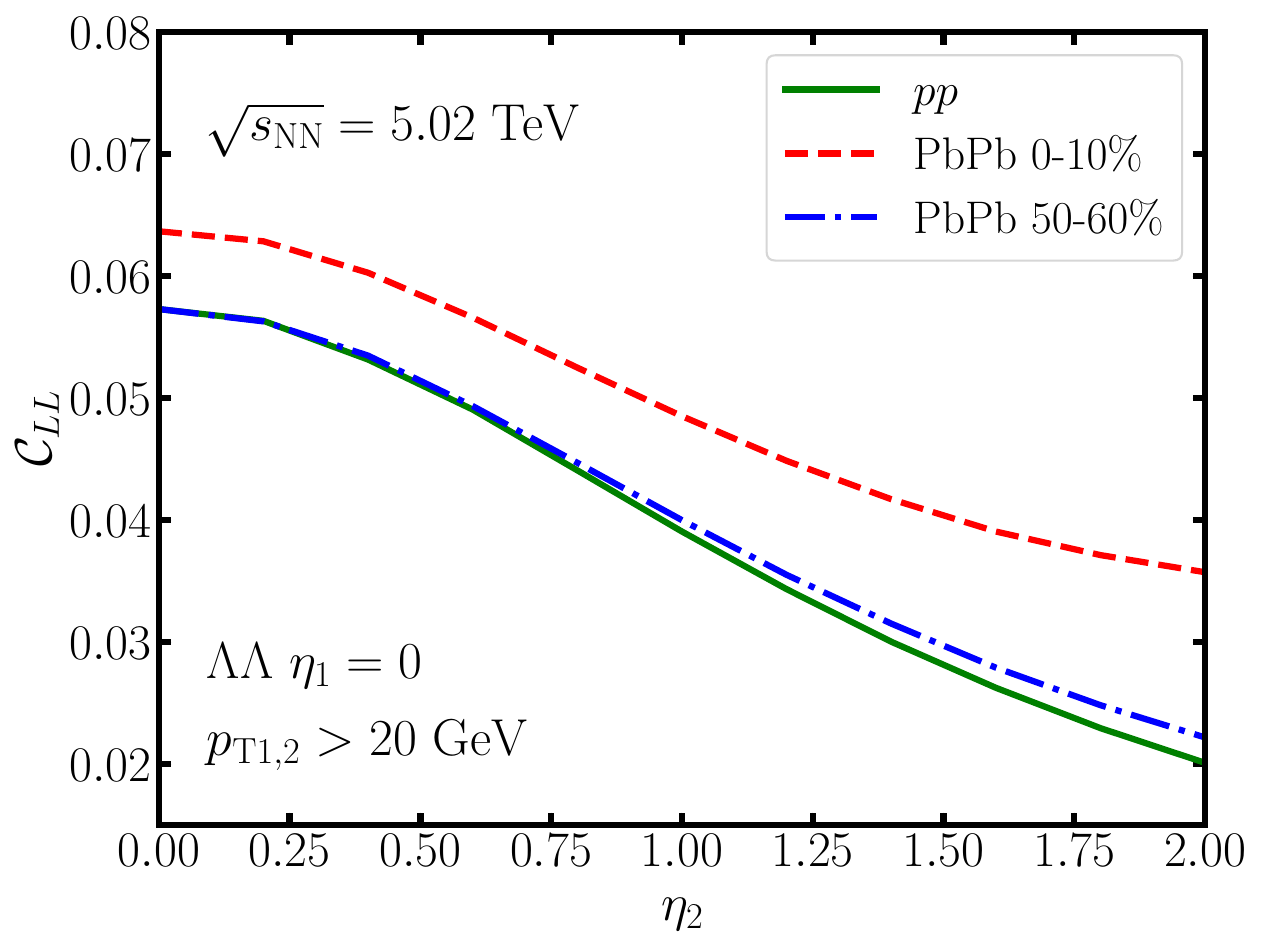}	
	\end{center}	
	\caption{(Color online) Correlation function of $\Lambda$-$\Lambda$ polarization at RHIC (left) and LHC (right) energies, compared between $pp$, peripheral AA and central AA collisions.}
	\label{AA3results}
\end{figure}

In Fig.~\ref{AA3results}, one can clearly observe an increase of $\mathcal{C}_\mathrm{LL}$ from $pp$ to peripheral AA and then to central AA collisions at both RHIC (left panel) and LHC (right panel). At mid-rapidity, $\mathcal{C}_\mathrm{LL}$ is enhanced by about 30\% from $pp$ to central Au+Au collisions at $\sqrt{s_\mathrm{NN}}=200$~GeV, and is enhanced by about 10\% from $pp$ to central Pb+Pb collisions at $\sqrt{s_\mathrm{NN}}=5.02$~TeV. Although on average, partons lose less energy in less energetic heavy-ion collisions at RHIC than at LHC, effect of energy loss on $\mathcal{C}_\mathrm{LL}$ appears stronger in the former system. This could be understood with the softer parton spectra produced at RHIC than at LHC, resulting in a larger shift of $z$ in FFs when energy loss is introduced. 

\begin{figure}[tbp!]
	\begin{center}
		\includegraphics[width=0.4\textwidth]{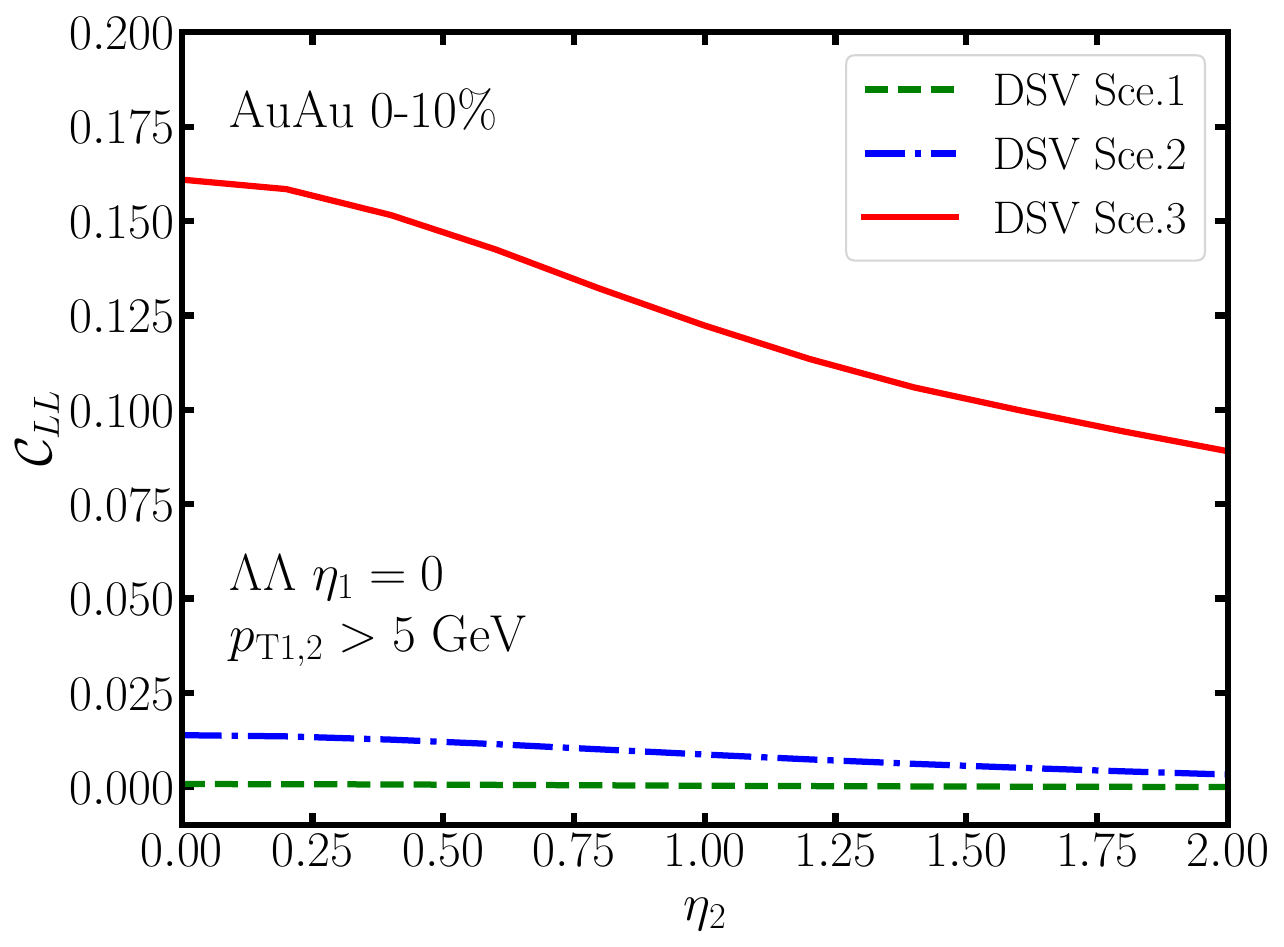}			
		\includegraphics[width=0.4\textwidth]{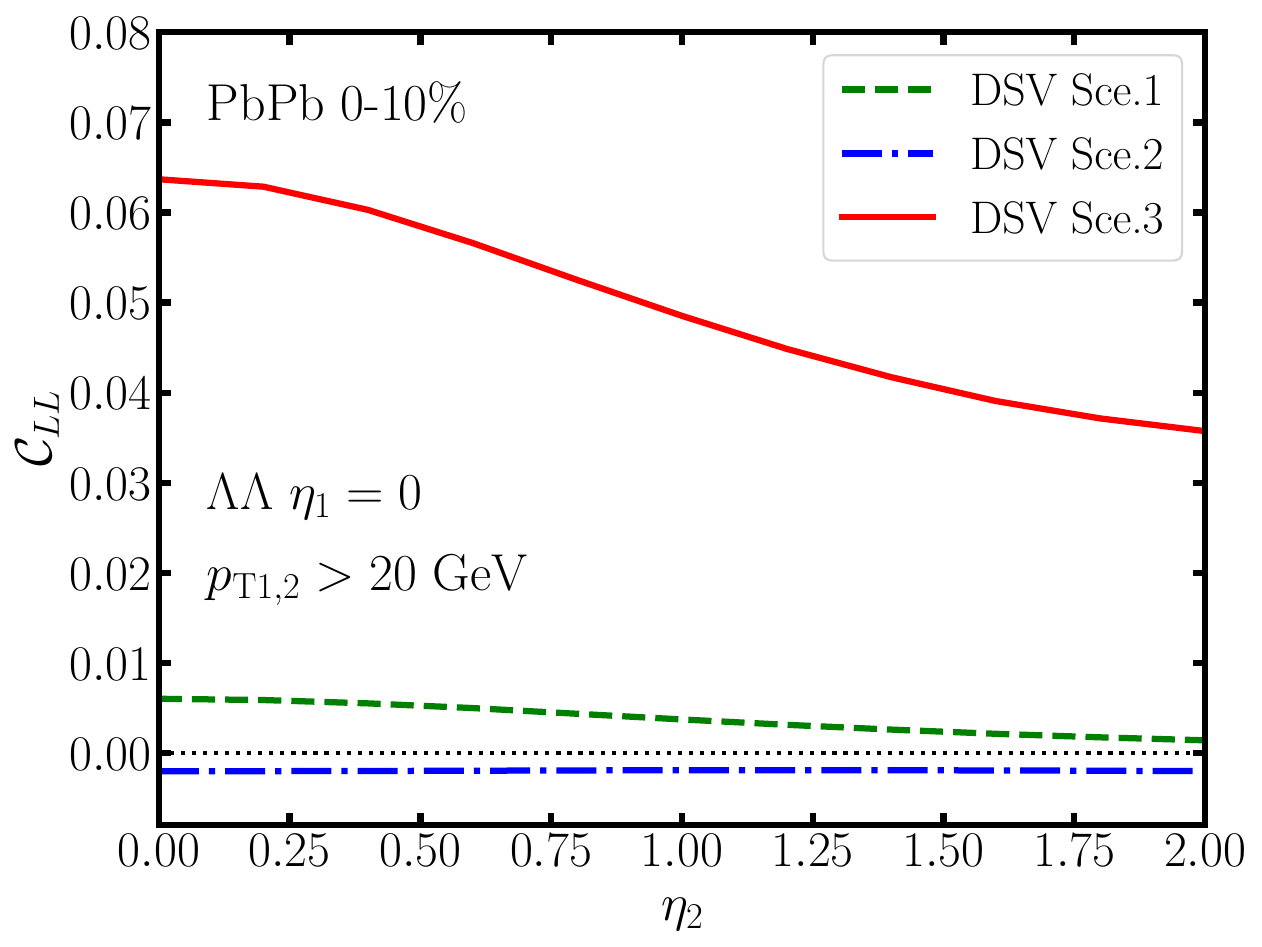}			
	\end{center}	
	\caption{(Color online) Correlation function of $\Lambda$-$\Lambda$ polarization in central Au+Au collisions at RHIC (left) and central Pb+Pb collisions at LHC (right), compared between different parameterizations of the DSV fragmentation functions.}
	\label{scenario}
\end{figure}

To investigate effects of FFs on the correlation of $\Lambda$-$\Lambda$ polarization, we compare results obtained using different parameterizations of the DSV FFs in Fig.~\ref{scenario}, left for central Au+Au collisions at RHIC, and right for central Pb+Pb collisions at LHC. The curves from Scenario 3 are the same as those in Figs.~\ref{AAjet} and~\ref{AA3results} with the same settings. Recall that while the unpolarized parts of these three scenarios are the same, only $s$ quarks contribute to the longitudinal spin transfer to $\Lambda$ in Scenario 1, negative transfer is introduced for $u$ and $d$ quarks in Scenario 2, and $u$, $d$, $s$ contribute equally to the spin transfer in Scenario 3. Therefore, Scenario 3 provides the strongest correlation of helicity between the two hyperons, while the other two provide much weaker correlations. Since the FFs depend on both the factorization scale and the fractional momentum $z$, whether Scenario 1 or 2 generates larger correlation may vary with the beam energy of nuclear collisions. The orders of the correlation functions from these three scenarios in AA collisions here are qualitatively consistent with those in $pp$ collisions shown in Ref.~\cite{Zhang:2023ugf}, although quantitative difference is present due to parton energy loss in AA collisions. 

\begin{figure}[tbp!]
	\begin{center}
		\includegraphics[width=0.4\textwidth]{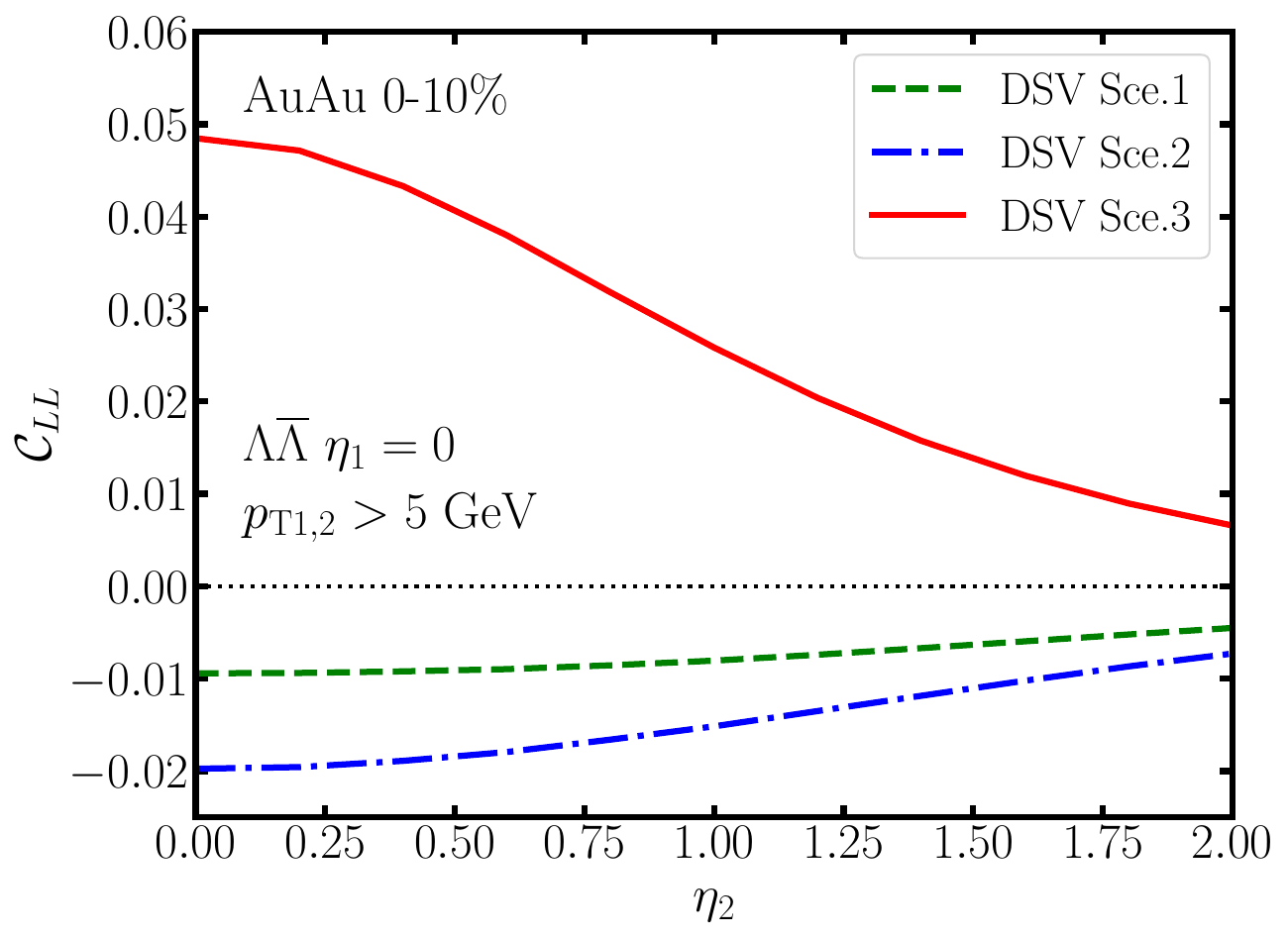}		
		\includegraphics[width=0.4\textwidth]{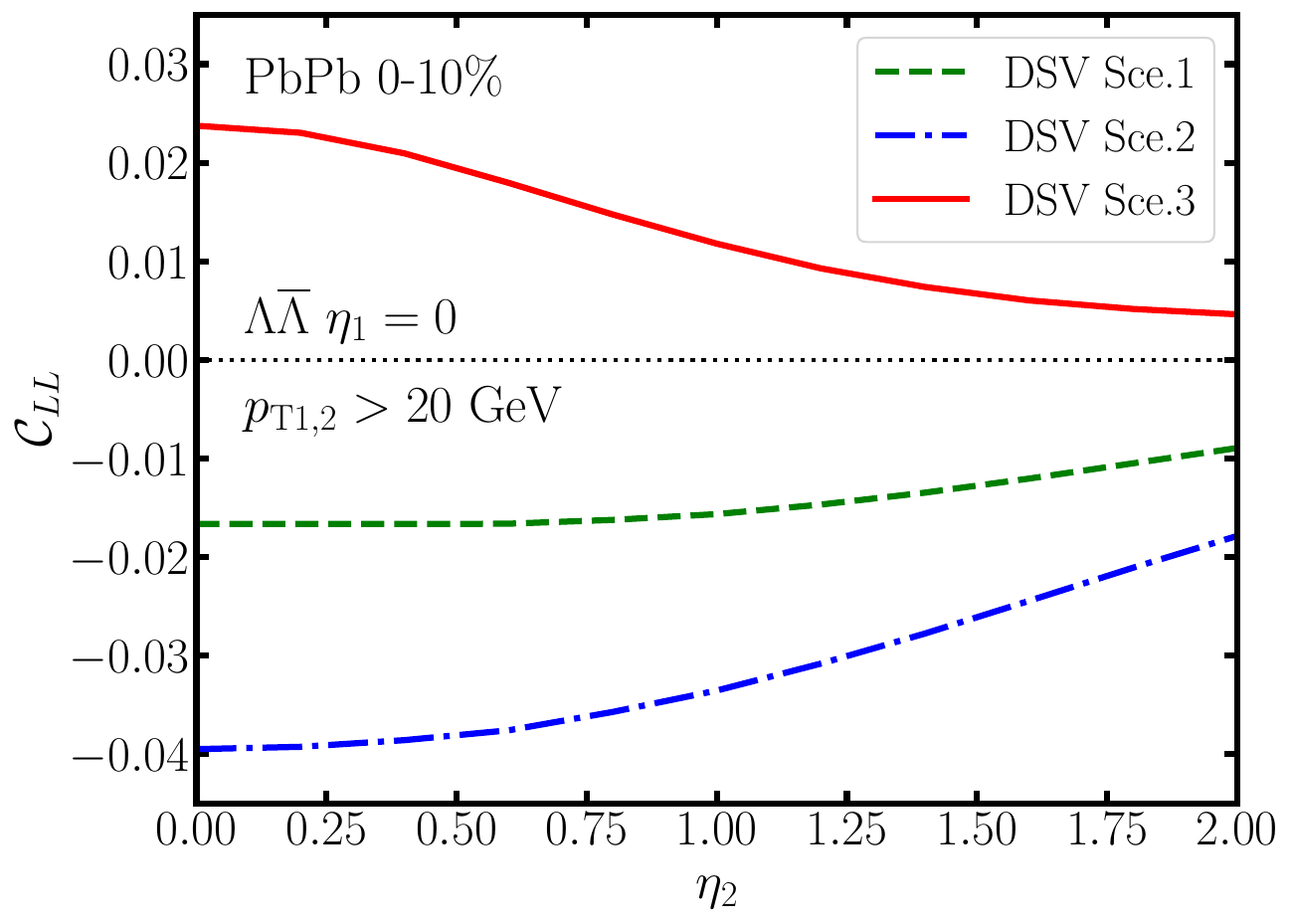}		
	\end{center}	
	\caption{(Color online) Correlation function of $\Lambda$-$\bar{\Lambda}$ polarization in central Au+Au collisions at RHIC (left) and central Pb+Pb collisions at LHC (right), compared between different parameterizations of the DSV fragmentation functions.}
	\label{Lambdabarscenario}
\end{figure}

In Fig.~\ref{Lambdabarscenario}, we further study the correlation function between $\Lambda$ and $\bar{\Lambda}$ in central Au+Au (left panel) and Pb+Pb (right panel) collisions. In Scenario 1, the major contribution to $\Lambda$-$\bar{\Lambda}$ production is through the $s$-channel creation of $s\bar{s}$ pair. This gives rise to the negative correlation of helicities between the two hyperons. In Scenario 2, although small negative $G_{1\rm L}/D_1$ is introduced from $u$ and $d$ quarks, the $G_{1\rm L}/D_1$ value for $s$ quarks becomes larger compared to that in Scenario 1~\cite{deFlorian:1997zj}. This can produce a stronger negative correlation function from Scenario 2 than from Scenario 1. In Scenario 3, the dominant contribution varies with the kinematics. As shown in Ref.~\cite{Zhang:2023ugf}, in $pp$ collisions, this correlation function is positive at small $\eta_2$ but negative at large $\eta_2$. After considering the parton energy loss, this correlation appears positive for Scenario 3 in both central Au+Au and central Pb+Pb collisions within the kinematic region we explore here. The different correlation functions given by different scenarios of the DSV FFs in Figs.~\ref{scenario} and~\ref{Lambdabarscenario} suggest this $\mathcal{C}_\mathrm{LL}$ could be a promising observable that improves our constraint on the spin dependent FFs. 

\begin{figure}[tbp!]
	\begin{center}
		\includegraphics[width=0.4\textwidth]{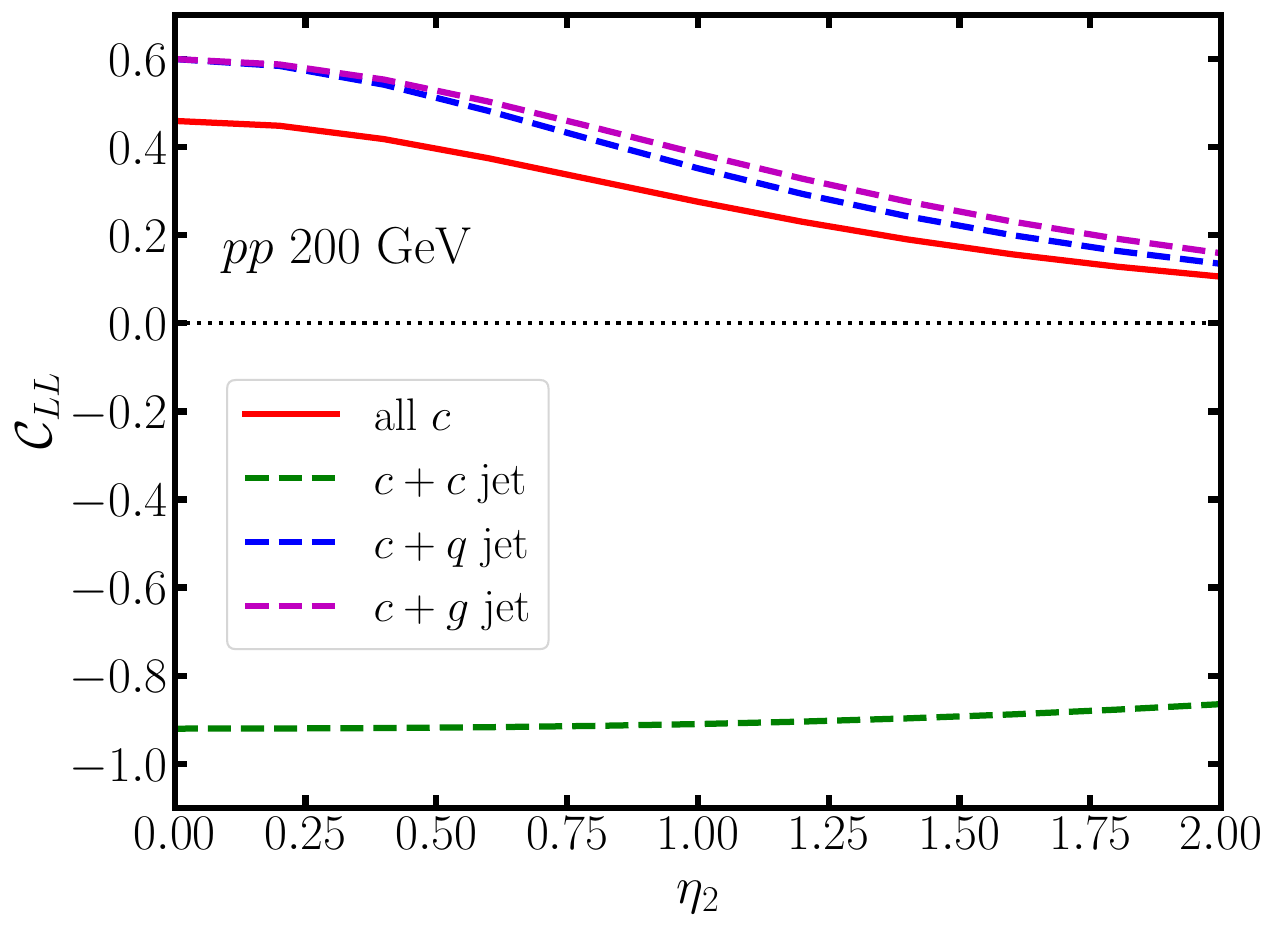}		
		\includegraphics[width=0.4\textwidth]{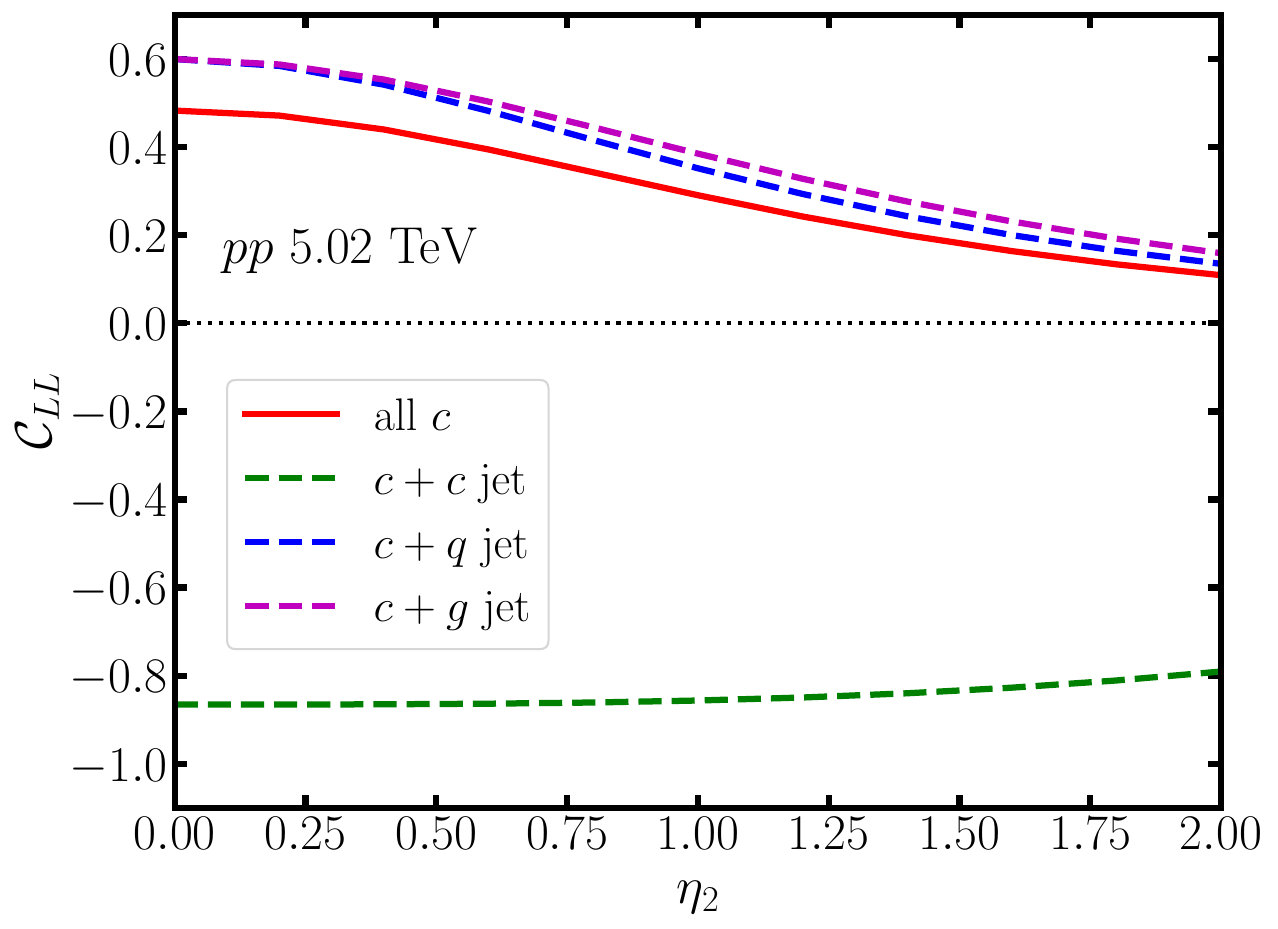}	
	\end{center}	
	\caption{(Color online) Helicity correlation between a $c$ quark and another parton produced at the LO in $pp$ collisions at 200~GeV (left) and 5.02~TeV (right).}
	\label{ppCjet}
\end{figure}

Due to our little knowledge on the spin-dependent FFs of $c$ quarks yet, we close this section by studying the spin correlation related to charm quarks on the partonic level. In Fig.~\ref{ppCjet}, we present the correlation function between a $c$ quark and another hard parton from the LO production in $pp$ collisions at 200~GeV (left panel) and 5.02~TeV (right panel). Here, we keep using the $5$~GeV cut on partons at the RHIC energy and $20$~GeV at the LHC energy. Because of the rarity of $c$ quarks inside the nuclear PDF, the production of a pair of $c$ quarks is dominated by the gluon fusion ($g+g \rightarrow c+\bar{c}$) and the quark annihilation ($q+\bar{q} \rightarrow c+\bar{c}$) processes, both of which are $s$-channel scatterings that conserve helicity. As a result, a strong negative correlation of helicity between the $c\bar{c}$ pair is expected. This might be observed through the $\Lambda_c$-$\bar{\Lambda}_c$ production in the future. On the other hand, the $c$-$g$ and $c$-$q$ pairs are produced though $t$-channel scatterings, whose correlation functions are positive. At the partonic level, little difference can be seen between our results at RHIC and LHC energies. And without convoluting with the FFs, we do not expect much difference introduced by parton energy loss in AA collisions either.

\section{Dihadron polarization in ultraperipheral heavy-ion collisions}
\label{section3}

Ultraperipheral nucleus-nucleus collisions (UPC) refer to a specific category of events where the impact parameter becomes so large that the overlapping region between the colliding nuclei disappears. Most likely, the two nuclei will pass each other without breaking up. Meanwhile, they will produce leptons and hadrons through the exchange of quasi-real photons ($\gamma^*$) or pomerons ($\mathbb{P}$). 

According to classical electrodynamics, a fast-moving nucleus is accompanied by enormous co-moving quasi-real
photons. However, the momenta of those quasi-real photons are not strictly collinear with that of the large nucleus. {The typical transverse momentum is about $1/R_{\rm{A}}$ with $R_{\rm{A}}$ the radius of the light source. For a large nucleus with $R_A \sim 7$ fm, the typical intrisic transverse momentum is around 30 MeV. Despite the small value, the intrinsic transverse momentum still gives birth to the dilepton azimuthal angular correlation \cite{Klein:2018fmp, Li:2019yzy, Li:2019sin, Klein:2020jom, Xiao:2020ddm, Shao:2022stc, Shao:2023zge} at the extremely back-to-back regime. In this process, the QCD multiple-soft-gluon radiation is absent. The photon intrinsic transverse momentum effect dominates at $0.99 \pi <\Delta \phi <\pi$ where $\Delta\phi$ is the azimuthal angle between two final state leptons. To properly take into account this effect, we shall resort to the 5D photon Wigner distribution function \cite{Klein:2020jom, Xiao:2020ddm}. On the other hand, for the QCD events, the parton shower effect results in a transverse momentum imbalance at around $10 \sim 30$ GeV which is much larger than the photon intrinsic transverse momentum. Therefore, the intrinsic transverse momentum of the incoming photon is negligible. Furthermore, in this work, the transverse momentum imbalance between two final state hadrons has been integrated over. Therefore, the photon intrinsic transverse momentum and the parton shower effect do not contribute. We thus can start with the following collinear photon distribution \cite{Iancu:2023lel} derived from the equivalent photon approximation \cite{Jackson:1998nia}}
\begin{equation}
x f( x) = \frac{2Z^2 \alpha}{\pi} \left[  \zeta K_0(\zeta)K_1(\zeta) - \frac{\zeta^2}{2} (k^2_1(\zeta) - K^2_0(\zeta)) \right] .
\end{equation}
Here, $\zeta = 2xM_{{p}} R_{\rm{A}}$ with $x$ the momentum fraction carried by the quasi-real photon, $M_{{p}}$ the proton mass and $R_{\rm{A}}$ the nucleus radius.

We consider the back-to-back dijet production in the UPC process. One of the jets produces a $\Lambda$ hyperon while the
other produces a $\bar{\Lambda}$. The kinematics are illustrated as follows
\begin{align}
\mathrm{AA} \rightarrow \mathrm{AA} + \Lambda(\eta_1,\lambda_1) + \bar{\Lambda}(\eta_2,\lambda_2) + X
\end{align}
where $\eta_{1(2)}$ and $\lambda_{1(2)}$ represent rapidity and helicity of the final state hadron. Back-to-back dijets in the UPC process are mainly produced through $\gamma \gamma$ and $\gamma \mathbb{P}$ interactions. We present our formulae respectively.

\begin{figure}[tbp!]
	\begin{center}
		\includegraphics[width=0.4\textwidth]{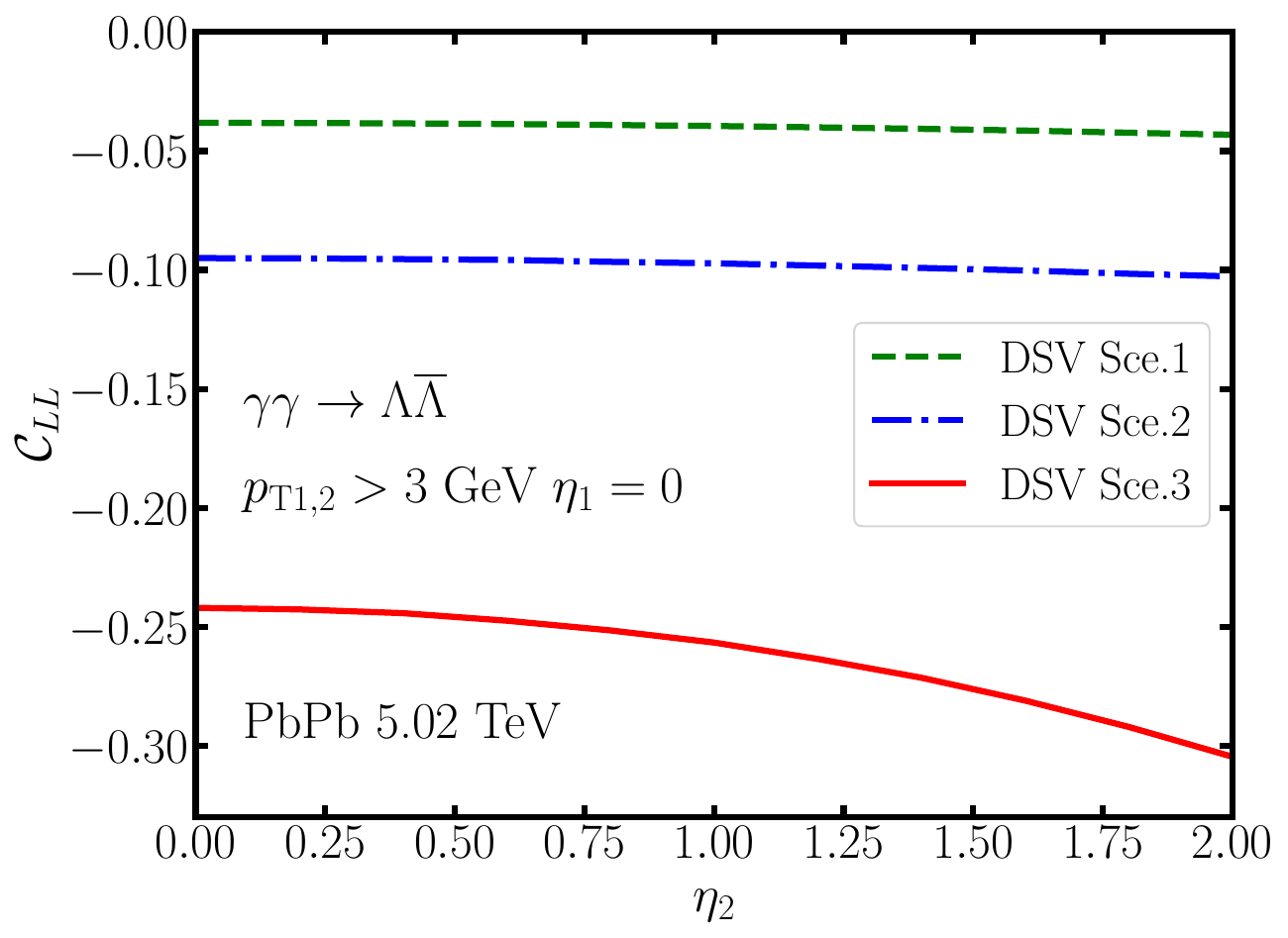}	
	\end{center}	
	\caption{(Color online) Correlation function of $\Lambda$-$\bar{\Lambda}$ polarization from the $\gamma\gamma\rightarrow \Lambda\bar{\Lambda}$ production in ultraperipheral Pb+Pb collisions at $\sqrt{s_\mathrm{NN}}=5.02$~TeV, compared between different parameterizations of the DSV fragmentation functions.}
	\label{gammaSenario}
\end{figure}

For the $\gamma \gamma$ channel, the differential cross section reads
\begin{align}
\frac{d\sigma^{\gamma \gamma}}{d\eta_1 d\eta_2} = \int dz_1 dz_2 \int \frac{d^2 \vec{p}_{\rm{T1}}}{z_1^2} x_1f_{\gamma}(x_1)x_2f_{\gamma}(x_2) \sum_q \frac{1}{\pi} \frac{d \hat{\sigma}^{\gamma \gamma \rightarrow q\bar{q}}}{dt} \left[ D^{\Lambda}_{1,q}(z_1) D^{\bar{\Lambda}}_{1,\bar{q}}(z_2) - \lambda_1 \lambda_2 G^{\Lambda}_{{1\rm L},q}(z_1) G^{\bar{\Lambda}}_{{1\rm L},\bar{q}}(z_2) \right],
\end{align}
where the exchange between $q$ and $\bar{q}$ is implicit here and the partonic cross section is
\begin{align}
\frac{d \hat{\sigma}^{\gamma \gamma \rightarrow q \bar{q}}}{dt} = \frac{6\pi \alpha^2 e^4_q}{s^2} \frac{t^2 + u^2}{tu}.
\end{align}
With the cross section above, one may evaluate the correlation function of dihadron polarization as defined in the first line of Eq.~(\ref{AAC}). Shown in Fig.~\ref{gammaSenario} is this correlation function from the $\gamma\gamma \rightarrow \Lambda \bar{\Lambda}$ process in ultraperipheral Pb+Pb collisions at $\sqrt{s_\mathrm{NN}}=5.02$~TeV, compared between different parameterizations of the DSV FFs. Here, both hyperons are required to possess transverse momentum greater than 3~GeV. Although this process resembles the $e^+e^-$ annihilation, the hard factor now reads $e^4_q$ instead of $e^2_q$. As a consequence, the production of $s\bar{s}$-pairs, relative to $u\bar{u}$, is strongly suppressed comparing to that in $e^+e^-$ annihilation. Recall that the three scenarios in the DSV parameterizations of $G_{{1\rm L},q}$ assume different flavor dependences. As shown in Fig.~\ref{gammaSenario}, the correlation from Scenario 3 is almost an order of magnitude larger than that from Scenario 1. This feature of the $\gamma \gamma$ channel may have a potential to further constrain which scenario neighbors reality. 

However, it is not plausible to isolate dijets produced though the $\gamma \gamma$ channel from the others. It has also been shown in Refs.~\cite{Iancu:2023lel,Iancu:2022lcw,Guzey:2016tek,Bertulani:1987tz} that the $\gamma \mathbb{P}$ channel accounts for almost 90\% of the dijet production. Moreover, it has been demonstrated that exclusive dijet production of $\gamma \mathbb{P} \rightarrow q \bar{q}$ is power suppressed~\cite{Iancu:2023lel,Iancu:2021rup,Iancu:2022lcw}. The dominant contribution actually arises from the (2+1)-diffractive process with an additional soft gluon emission. For simplicity, we always call the photon-going direction the positive rapidity direction. The exchange between $\gamma$ and $\mathbb{P}$ is implicit as well. Taking the soft gluon limit and integrating over the redundant phase space, the differential cross section now reads
\begin{equation}
\begin{aligned}
\frac{d\sigma^{\gamma \mathbb{P}}}{d\eta_1 d\eta_2} = &\int \frac{dx}{x} \int dz_1 dz_2 \int \frac{d^2 \vec{p}_{\rm{T1}}}{z_1^2} x_1f_{\gamma}(x_1)xg_{\mathbb{P}}(x,Q^2) \sum_q \frac{1}{\pi} \frac{d \hat{\sigma}^{\gamma \mathbb{P} \rightarrow q\bar{q}}}{dt} \\
&\times \left[ D^{\Lambda}_{1,q}(z_1) D^{\bar{\Lambda}}_{1,\bar{q}}(z_2) - \lambda_1 \lambda_2 G^{\Lambda}_{{1\rm L},q}(z_1) G^{\bar{\Lambda}}_{{1\rm L},\bar{q}}(z_2) \right],
\end{aligned}
\end{equation}
where the phase space for the $x$-integral is constrained by the following requirement
\begin{align}
	|\eta_g| = \left| \ln{\frac{p_{\rm{T1}}}{z_1K^{\rm{typical}}_{\perp}}} + \ln(e^{-\eta_1}+e^{-\eta_2}) + \ln{\frac{1-x}{x}} \right| < 4.5, 
\end{align}
with $K^{\rm{typical}}_{\perp} = 2$ GeV. The integrated gluon distribution function of the pomeron $xg_{\mathbb{P}}(x,Q^2)$ is
\begin{align}
xg_{\mathbb{P}}(x,Q^2) = \frac{S_{\perp} (N_c^2-1)}{4\pi^3} \frac{1}{2\pi (1-x)} \int_0^{Q^2} dK^2_{\perp} \left[ \mathcal{M}^2 \int _0^{\infty} dR R J_2(K_{\perp} R) K_2(\mathcal{M}R) { \mathcal{T} }_g(R) \right]^2 ,
\end{align}
with $Q$ playing the role of the factorization scale, $\mathcal{M}^2 = xK^2_{\perp}/(1 - x)$, $S_{\perp}$ is of constant with dimension of GeV$^{ -2}$ and $\mathcal{T}_g(R)$ the dipole scattering amplitude in the adjoint representation. In the McLerran-Venugopalan (MV) model~\cite{McLerran:1993ni,McLerran:1994vd}, $\mathcal{T}_g(R)$ is given by
\begin{align}
\mathcal{T}_g(R) = 1-\exp \left[ -\frac{1}{2} Q^2_{\rm{A}} R^2 \ln{\left( e+\frac{2}{\Lambda R} \right)} \right],
\end{align}
with $Q_{\rm{A}} = 0.303$ GeV$^2$ and $\Lambda  = 0.2$ GeV. Taking the soft gluon limit, it is straightforward to obtain the partonic cross section from Ref.~\cite{Iancu:2023lel} as
\begin{align}
\frac{d \hat{\sigma}^{\gamma \mathbb{P} \rightarrow q \bar{q}}}{dt} = \frac{\pi \alpha \alpha_\mathrm{s} e^2_q}{s^2} \frac{t^2 + u^2}{tu}.
\end{align}

\begin{figure}[tbp!]
\begin{center}
\includegraphics[width=0.4\textwidth]{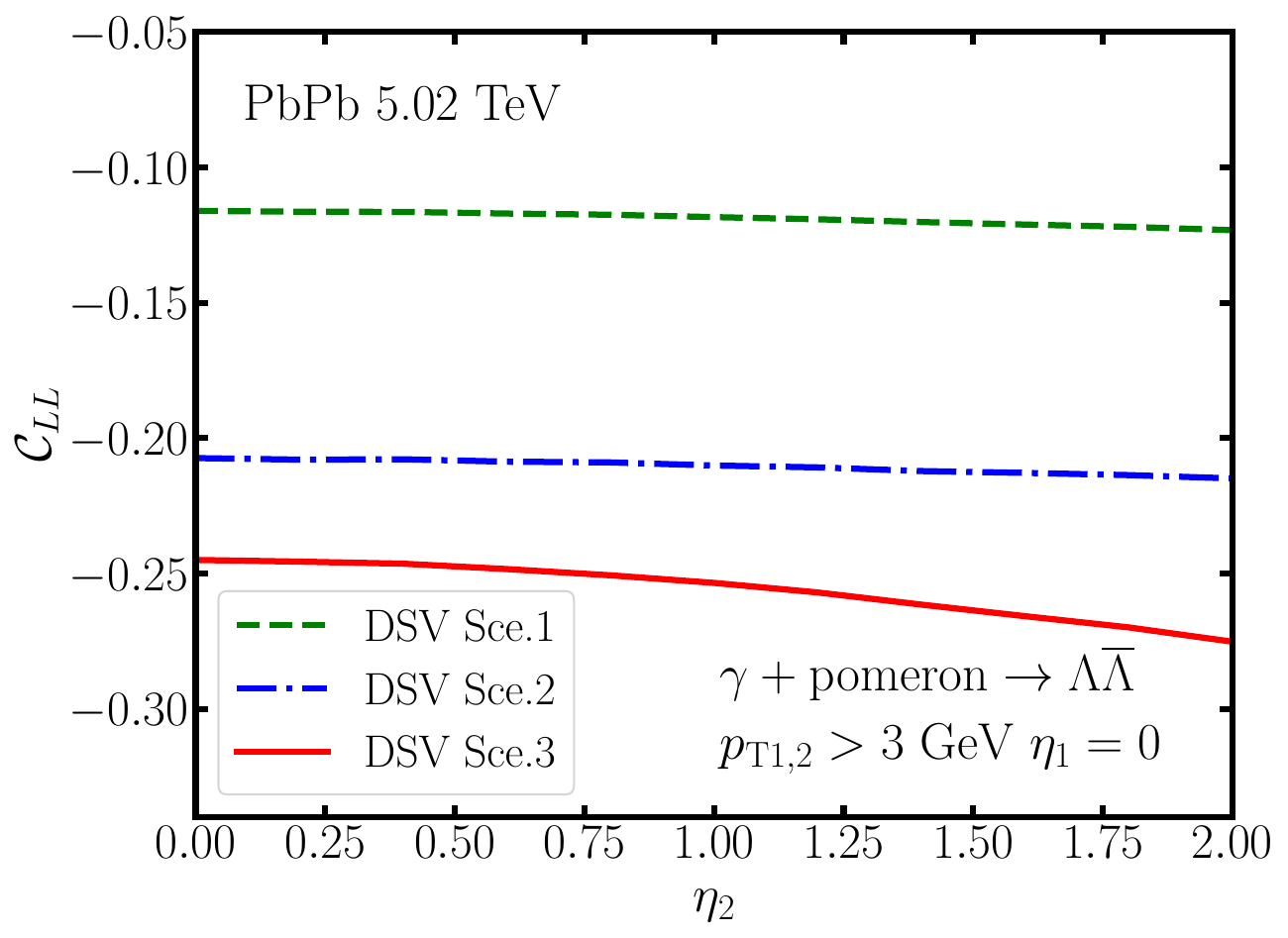}		
\end{center}	
\caption{(Color online) Correlation function of $\Lambda$-$\bar{\Lambda}$ polarization from the $\gamma \mathbb{P} \rightarrow \Lambda \bar{\Lambda}$ production in ultraperipheral Pb+Pb collisions at $\sqrt{s_\mathrm{NN}}=5.02$~TeV, compared between different parameterizations of the DSV fragmentation functions.}
\label{pomeron}
\end{figure}

Apparently, the flavor dependent part of the hard factor becomes $e^2_q$ again. Since this is the dominant channel, we expect that the correlation of dihadron polarization in UPC exhibits similar features to those in the $e^+e^-$ annihilation process. Indeed, as shown in Fig.~\ref{pomeron}, for the $\gamma\mathbb{P} \rightarrow \Lambda \bar{\Lambda}$ process, the correlation functions obtained from different scenarios of the DSV FFs are on the same order of magnitude, similar to the $e^+e^-$ process~\cite{Zhang:2023ugf}.

\section{summary}
\label{section4}

We investigate the correlation of dihadron polarization in high-energy nuclear collisions. By convoluting the spin-dependent cross section of hard parton production, the parton energy loss inside the QGP, and the spin-dependent fragmentation function, we evaluate the correlation function of helicity ($\mathcal{C}_\mathrm{LL}$) between a pair of $\Lambda$ ($\bar{\Lambda}$) hyperons at the leading order and the leading twist, and explore its variation from $pp$ collisions to peripheral and central heavy-ion collisions. Within the third scenario of the DSV FFs, where $u$, $d$ and $s$ quarks contribute equally to $\Lambda$, we find a sizable enhancement of $\mathcal{C}_\mathrm{LL}$ between $\Lambda$-$\Lambda$ pairs from  $pp$ collisions to central AA collisions. This is mainly due to the parton energy loss that allows the final $\Lambda$ within a given $p_\mathrm{T}$ range to probe a larger $z$ regime of the FFs, where the longitudinal spin transfer $G_{1\rm L}/D_1$ is stronger. The value of $\mathcal{C}_\mathrm{LL}$ at the RHIC energy is larger than that at the LHC energy, because of the higher $z$ probed at the former. Different scenarios of the DSV FFs are shown to generate significantly different $\mathcal{C}_\mathrm{LL}$ between $\Lambda$-$\Lambda$ pairs. Considering the higher luminosity of jet production and the enhanced value of the helicity correlation in AA collisions than in $pp$ collisions, relativistic heavy-ion collisions may provide a novel opportunity to constrain the polarized FFs without requiring polarized beams. Similar sensitivity on the longitudinal spin transfer is also shown for the $\Lambda$-$\bar{\Lambda}$ polarization in heavy-ion collisions at RHIC and LHC. At the partonic level, a strong negative correlation, close to -1, is found between a $c\bar{c}$ pair produced in energetic nuclear collisions. Since $c$ quarks evolve through the QGP with their flavor conserved, this may become a clean channel for probing the polarized parts of the FFs.

The correlation of dihadron polarization is also explored for ultraperipheral heavy-ion collisions, where two major processes of hyperon production, $\gamma \gamma \rightarrow \Lambda \bar{\Lambda}$ and $\gamma \mathbb{P} \rightarrow \Lambda \bar{\Lambda}$, are studied separately. It is interesting to note that since the partonic hard cross section of the former process is proportional to $e_q^4$ instead of $e_q^2$, an order of magnitude difference is seen in the $\mathcal{C}_\mathrm{LL}$ function between the three scenarios of the DSV FFs. However, because the $\Lambda$-$\bar{\Lambda}$ production is dominated by the latter process, whose hard cross section is proportional to $e_q^2$, a similar $\mathcal{C}_\mathrm{LL}$ is expected in UPC to that in the $e^+e^-$ annihilation process. Considering the relatively higher $p_\mathrm{T}$ of jets here compared to that obtained in the current $e^+e^-$ programs, ultraperipheral heavy-ion collisions provide a complementary means to improve our knowledge on the polarized FFs. Another major area of UPC physics is the photoproduction of vector mesons~\cite{Hagiwara:2020juc,Xing:2020hwh,Zha:2020cst,Brandenburg:2022jgr,Wu:2022exl}. It has been recognized that linearly polarized photons can give rise to an azimuthal asymmetry in the vector meson distribution in UPC, which encodes rich information about the gluonic matter distribution inside nucleus and is manifested in a unique spin interference pattern seen in experiments~\cite{STAR:2022wfe}. Therefore, it would also be interesting to relate our current study on the polarization of hyperons to their azimuthal distributions in the future.

\section*{Acknowledgments}
We thank Jian Zhou and Wen-Jing Xing for fruitful discussions and helpful suggestions. This work was supported by the National Natural Science Foundation of China (NSFC) under Grant Nos.~12175122 and 2021-867. S.Y.~Wei is supported by the Taishan fellowship of Shandong Province for junior scientists and the Shandong Province Natural Science Foundation under grant No.~2023HWYQ-011.

\bibliographystyle{h-physrev5}
\bibliography{ref}
\end{document}